\documentclass{CUP-JNL-DCE}

\usepackage{latexsym}
\usepackage{graphicx}
\usepackage{multicol,multirow}
\usepackage{amsmath,amssymb,amsfonts}
\usepackage{mathrsfs}
\usepackage{amsthm}
\usepackage{apacite}
\usepackage{rotating}
\usepackage{appendix}
\usepackage[authoryear]{natbib}
\usepackage{ifpdf}
\usepackage[T1]{fontenc}
\usepackage{times}
\usepackage{sourcesanspro}
\usepackage{newtxmath}
\usepackage{textcomp}%
\usepackage{xcolor}%
\usepackage{hyperref}
\usepackage{cleveref}
\crefformat{footnote}{#2\footnotemark[#1]#3}
\usepackage{todonotes}
\usepackage{tabularx}
\usepackage{placeins}
\usepackage[frozencache=true,cachedir=minted-cache]{minted}
\setminted[cypher]{
    breaklines=true,
    encoding=utf8,
    fontsize=\footnotesize,
    frame=lines
}
\usepackage{etoolbox}

\makeatletter
\AtBeginEnvironment{minted}{\dontdofcolorbox}
\def\dontdofcolorbox{\renewcommand\fcolorbox[4][]{##4}}
\makeatother

\articletype{RESEARCH ARTICLE}
\jname{Data-Centric Engineering}
\jyear{2022}
\jdoi{10.1017/dce.2020.7}

\begin{document}

\begin{Frontmatter}

\title{An Approach for System Analysis with MBSE and Graph Data Engineering}

\author*[1]{Florian Schummer}\email{f.schummer@tum.de}
\author[1]{Maximillian Hyba}\email{}

\authormark{Florian Schummer\textit{et al.}}

\address[1]{\orgdiv{Chair of Astronautics}, \orgname{Technical University of Munich}, \orgaddress{\street{Boltzmannstr 15}, \state{Bavaria}, \postcode{Garching, 85748}, \country{Germany}}}

\received{TBD}
\revised{}
\accepted{}


\abstract{
Model-Based Systems Engineering aims at creating a model of a system under development, covering the complete system with a level of detail that allows to define and understand its behavior and enables to define any interface and workpackage based on the model.
Once such a model is established, further benefits can be reaped, such as the analysis of complex technical correlations within the system.
Various insights can be gained by displaying the model as a formal graph and querying it.
To enable such queries, a graph schema needs to be designed, which allows to transfer the model into a graph database.
In the course of this paper, we discuss the design of a graph schema and MBSE modelling approach, enabling deep going system analysis and anomaly resolution in complex embedded systems.
The schema and modelling approach are designed to answer questions such as what happens if there is an electrical short in a component? Which other components are now offline and which data cannot be gathered anymore? Or if a condition cannot be met, which alternative routes can be established to reach a certain state of the system.
We build on the use case of qualification and operations of a small spacecraft.
Structural and behavioral elements of the MBSE model are transferred to a graph database where analyses are conducted on the system.
The schema is implemented by an adapter for MagicDraw to Neo4j.
A selection of complex analyses are shown on the example of the MOVE-II space mission.

}
\begin{policy}[Impact Statement]
The proposed schema to transfer SysML Models to Labelled Property Graphs and the related modelling strategy open a wide range of system analyses for designers and operators of complex embedded systems, such as tracking data paths, finding probable causes of anomalies or retrieving every possibility to reach a certain state within the system from the current one.
It can also be used to analyze which input losses components should be resilient against to achieve a maximum level of robustness and may help modelers in choosing what information to include in their model.
By parametrizing graph-queries, they can be reused for any system and thereby increase the efficiency of model analysis.
The requirements imposed on the analyzed SysML model are few and do not require a holistically complete model.
Thereby the approach can also be applied to system models still in the making or where parts of the system are black-boxed.
\end{policy}

\end{Frontmatter}

\section{Introduction}
\label{sec1}

\subsection{Obstacles in the Introduction of Model Based Systems Engineering}
In 2007, the International Council on Systems Engineering defined Model Based Systems Engineering as
"the formalized application of modeling to
support system requirements, design, analysis, verification and validation activities beginning in the conceptual design phase and continuing throughout development and later life cycle
phases."
\cite[p. 15]{incose-strategy-2007}

In 2019, the National Aeronautics and Space Administration published a survey with 50 participants from the industry, academia, US-government agencies and tool vendors for systems engineering tools on the topic of the current state of systems engineering at their own institution or as perceived at partnering entities (cf. \cite{nasa-se-study2019}).
The study identified an increased application of Model Based Systems Engineering (MBSE) as the number 3 factor on improving overall performance, after improving training of systems engineers and improving their domain specific knowledge.
Asked about the expected benefits,  63\% expect a reduction of 30\% to 50\% regarding the development cycle time.
While encouraging the use of MBSE, the study also found the majority of participants reporting less than 25\% adoption of MBSE in projects at their workplace.
33\% of the participants across all disciplines see insufficiency of MBSE tools as a key weakness in their application of systems engineering.
The only factor named more often are cultural issues with 46\%, which the authors describe as people being reluctant to change from document based processes (\cite{nasa-se-study2019}).

I chose this study for the introduction to this paper for the following reasons:
According to the study, systems engineers across industry, public agencies and academia show a high level of expectations on MBSE.
At the same time, they admit that the tools are inadequate.
People do not accept tools that they view as inferior to their current working environment, thus leading to a smaller percentage of adaption of MBSE, than the expected improvement through its application would suggest.

The successful introduction of new products and processes always requires understanding the users' needs.
Roughly around the same time as NASA's study was conducted, the European Space Agency started its "Model Based for Systems Engineering" initiative, which aims at increasing the usability and interoperability of MBSE and MBSE tools within the European space sector.
A report on perceived user needs published by the initiative in 2020 sums up the current state of interoperability between tools as low, resulting in unnecessary duplicate work.
The report identifies 15 key user needs that MBSE should deliver.
One of the key user needs is to "ensure the consistency, completeness and feasibility of requirements and design", noting that "especially the functional complexity must be kept in [sic] control"\cite[p. 11]{mb4se-2020-userneeds}.
Another key user need is to "structure the knowledge about the system in such a way that the relations between the knowledge elements are established and traceable"\cite[p. 11]{mb4se-2020-userneeds}.

Looking at yet another survey on the adoption of MBSE, the report \textit{Benchmarking the Benefits and Current Maturity of Model-Based Systems Engineering across the Enterprise} by the Systems Engineering Research Center asked 240 individuals across academia and industry what they see as the largest obstacles in MBSE adoption.
"MBSE methods and processes" is cited as number one obstacle by the report \cite[p.20]{serc2020}.
\cite{nasa-se-study2019} supports these findings;
88\% of participants identified "Purpose and Scope Definition" as a major challenge in the adoption of MBSE.

Combining the results of the three reports gives insight into some obstacles to rolling out MBSE on a wider scale;
MBSE tools shall enable users to ensure the completeness, consistency and feasibility of their development and a third of the participants in NASA's study identify insufficiency of MBSE tools as a key weakness to their application of MBSE.
The conclusion lies near, that the tools are not yet fully up to the task.
Combined with the findings of \cite{serc2020}, the reports show tools and modelling strategies as major factor for a broader acceptance of MBSE.
Part of the tools inadequacy in our experience is their inability to process complex queries on models.
Starting at any given element of a model, one can usually only follow it for one hop, i.e. to the end of any of its direct relations, or along a single relation type.
Although any consistent model contains complex modelling patterns, these can usually not be followed within standard commercial tools.

\subsection{Goals and Outline}


This paper shall show how graph databases can be used for specific MBSE analyses that are cumbersome and inefficient in common modelling tools such as MagicDraw.
In the course of this document typical analysis tasks during spacecraft development, formalized as questions shall be outlined.
Furthermore it shall be explored how graph-technology enables such analyses.
A graph schema is a prerequisite for any graph-based analysis.
A graph schema transforming SysML into a labelled property graph shall hence be developed and a respective software implementation, which automatically transfers standard SysML models into the graph-database shall be provided.
A goal of this paper is to enable others to conduct such analyses as well.
Therefore a complementary set of modelling guidelines shall be presented, which is to be kept to a bare minimum to reach a high compatibility with any prevailing modelling strategy.
Customizations of the modelling guidelines shall be discussed, depending on which analyses are to be carried out.
Finally, a selection of the previously defined analysis tasks shall be translated into actual queries for the graph database and performed on the example of an actual spacecraft.
Before getting into the details of analysis tasks, graph schemata and modelling rules, some background shall be provided.

Therefore, \Cref{sec:stateoftheart} gives an overview of various graph databases after summarizing the current state of MBSE implementation as gathered in recent surveys.
It concludes with a short introduction to the use case of the MOVE-II spacecraft.
\Cref{sec:Development of the Graph Schema} outlines typical tasks in system analysis, with a special focus on later project phases such as Assembly Integration Testing and Operations of a system.
Building on the outlined tasks, possible solutions for a graph schema are discussed and a solution suitable to the analysis tasks is selected.
In \Cref{sec_modelling_guideline} modelling guidelines for the MBSE model are presented and their effect on which analyses can be conducted is discussed.
\Cref{sec_implementation} provides a short overview of the actual code implementation, which translates MagicDraw SysML models into a labelled property graph.
In \Cref{sec_application} the proposed graph schema and modelling guidelines are verified against an operational space mission:
The MOVE-II space mission designed and operated by the Technical University of Munich's Chair of Astronautics.
The model used for the analysis covers the complete structure of the spacecraft, including hardware and software, as well as any systems and components of the ground station and operations equipment.
\Cref{sec_conclusion} compares the results of the paper to the state of the art and \Cref{sec_outlook} presents an outlook on future topics of research in the area.

\FloatBarrier
\section{Background}
\label{sec:stateoftheart}
This section is divided in three parts;
The first part covers the state of the art on the implementation of MBSE and the challenges thereof as perceived by the literature.
No previous work that compares different studies on the implementation of MBSE could be found.
Also, no previous work was found on the application of MBSE in later phases of a spacecraft project such as Assembly Integration and Testing or Operations of the spacecraft.
The comparison is conducted here to provide a better impression on what challenges people are dealing with when implementing MBSE and to gain insight into why MBSE seems to not be applied to later life cycle phases.
The second part introduces graph databases and the underlying concept of graph schemata.
The third part gives a short introduction on the MOVE-II project used as implementation show cases.

\subsection{Recent Surveys on Model-Based Systems Engineering}
\label{ssec:BGMBSEsurveys}
The report \textit{Benchmarking the Benefits and Current Maturity of Model-Based Systems Engineering across the Enterprise} by \cite{serc2020} was published in March 2020.
Studies lying to far in the past are seen as less relevant for this work, as the presented studies show, that the application of MBSE drastically increased over the in recent times and new challenges and problems arose while others become less relevant.
Being supported by the United States' Department of Defense, the report sets out to explore the current state of adoption of MBSE, identifying challenges and enablers in the adoption and which skillsets are necessary for a successful MBSE adoption.
The survey is based on data gathered from November 2019 to January 2020, with a total of 240 participants and thereby comprises not only the most recent but also largest survey, with participants from industry, academia and governmental institutions.
While \cite{serc2020} and \cite{nasa-se-study2019} were the most recent surveys we could find, other publications address the same topic and should not be omitted.

\cite{chamimbsesurvey} report on a survey conducted on MBSE adoption in the industry with 42 participants.
The survey is focused on MBSE adoption challenges.
In difference to \cite{nasa-se-study2019}, all participants had an industrial background.
The participants came from a broader field, with roughly a third working in consultancy and training, and the rest being split on aerospace, medical, railway, defense, computing and IT engineering and automotive (ordered by decreasing participation).
The main findings are consistent with \cite{nasa-se-study2019}: "Awareness and change resistance" is the most frequently named challenge with 88\% of participants either agreeing or strongly agreeing on the topic.
Interesting is that "purpose and scope definition" has the same level of total agreement (88\%) but a lower percentage of strongly agreeing participants.
The third and fourth most mentioned problems are "method definition and extension" (84\%) and "tool dependency and integration" (83\%).

The higher percentages in agreement can be traced to the surveying method.
While \cite{nasa-se-study2019} built upon telephone interviews with rather free questions and consolidated the responses afterwards into groups, \cite{chamimbsesurvey} used a predefined set of perceived challenges in MBSE adoption and asked the survey participants about their agreement with each challenge.
However, it should be mentioned that "method definition and extension", as well as "purpose and scope definition" came up as additional challenges that were not in the focus of \cite{nasa-se-study2019}.
This can either be attributed to the broader field of participants or the different surveying method, which might bring up focus on topics that do not immediately spring to mind otherwise.

\cite{morrismbseissues} focused on the earlier phases of a system's lifecycle by limiting their two surveys from 2015 and 2014 to conceptual design works.
What makes the survey especially worth reading is the open-minded approach to the question whether the employment of MBSE brought benefits or exacerbated existing problems further.
Interestingly, the study found "solutioneering" (which is described as stakeholders pressing for a specific solution without understanding the problem first) and "lack of Stakeholder engagement" as main topics.
Topics mentioned in \cite{nasa-se-study2019} and \cite{chamimbsesurvey} such as "purpose and scope definition", "method definition and extension" or "tool dependency and integration" are not present in the report of Morris et al.
Likely reasons for the difference between the former two studies and \cite{morrismbseissues} might be the limited number of participants in any of the surveys, the 4-5 years difference between their conduction or the different methodologies and specific questions employed by the survey conductors.
\Cref{tab:mbsesurveys} gives a short overview of key numbers of all four publications.

\begin{table}[!t]
  \caption{Comparison of surveys on the state of Model Based Systems Engineering}
  \label{tab:mbsesurveys}
  \begin{tabularx}{\textwidth}{|l|X|l|l|X|}
    \hline
    \textbf{Year} & \textbf{Title} & \textbf{Authors} &  \textbf{No. Part.} & \textbf{Method} \\\hline
    2014/2015 & Issues in conceptual design and mbse successes: Insights from the model-based conceptual design surveys       & B. Morris et al.    & 39 / 40  & Open Questions      \\\hline
    2018 & A Survey on MBSE Adoption Challenges & M. Chami, J. Bruel &   42 & Level of agreement to preselected statements \\\hline
    2019 & Independent Assessment of Perception From External/non-NASA Systems Engineering (SE) Sources &  G. Pawlikowski et al. &  50 & Open questions \\\hline

    2020 & Benchmarking the Benefits and Current Maturity of Model-Based Systems Engineering across the Enterprise & T. McDermott et al. &  240 & combined \\\hline
  \end{tabularx}
\end{table}

Going on from the topic of MBSE challenges to specific lacks of functionality for MBSE tools, \cite{hazle2020} summarize literature on the verification and validation of SysML models.
They specifically point out reviews, analysis and simulation as the three main techniques of verification and validation for SysML models and describe formal model checking of SysML models as rarely used and often reserved for high risk aspects due to the significant effort attributed to it.
They describe the necessity to transform the model into another formal language prior to formal analyses, such as Petri Nets or directed graphs.
They further state the formal verification of models with Petri Nets or directed graphs is limited to the behavior aspects of a SysML model, omitting the structural aspects.
As for the analysis of requirements and structural aspects of the model, a wide variety of publications were made on the topic of automatic analysis of  requirements in SysML (see \cite{bankauskaite2018,petnga2019,morkevicius2015}).

What all presented publications have in common is their focus on the early project phases and formal reviews.
Going back to the original definition of MBSE, later life cycle phases such as the qualification and operation of a system should not be omitted but are rarely studied.
I found only one publication that can be linked to  MBSE employment in qualification and operation of a spacecraft.
In 2020, the European Space Agency issued an invitation to tender on reverse-engineering of the satellite OPS-SAT: The satellite is built as a generic flying laboratory, that allows new experiments to be uploaded and conducted while in orbit. The invitation to tender states the following intent:
"Users could greatly benefit from understanding of the system via a formal system model also for experiment integration and feasibility analysis" \cite[p.6]{opssatmbseitt}.
While the aim of this activity is the employment of MBSE to assist in the operational phase, the activity is ongoing and no publications on their progress are issued yet.
The lack of publications on MBSE in the operational phase and during assembly integration and testing leads to the conclusion that activities in this  field of application are rare.
Taking a look at the current rate of MBSE employment this is no surprise.
\cite{nasa-se-study2019} found a majority of interview partners reporting less than 25\% of MBSE on current projects, with a peak reporting between 5\% and 10\%.
MBSE is easier to employ in earlier project phases, where the system's design still has a lower level of detail and hence requires a lower degree of complexity from the model.
The conclusion lies near that due to the overall low level of adaption the focus lies on the earlier project phases.
Additionally it requires a detailed model of the spacecraft during Assembly Integration and Testing to reap any benefits of MBSE for AIT activities.
Such a detailed model implies that the project successfully implemented MBSE throughout the project's timeline, including any subcontractors.
As protection of intelectual property is a concern in spacecraft developments, subcontractors are naturally reluctant to provide a detailed model with their subsystems.
For the application of MBSE in Assembly Integration and Testing as well as Operations this leads to three conclusions:
\begin{enumerate}
  \item Any modelling guidelines built here to enable analyses via graph databases should be compliant with pre-existing modelling strategies, as ideally the model is maintained since early project phases and has a pre-existing modelling strategy
  \item In case no previous model exists and MBSE and graph analyses shall be employed in a later project phase, building the model should be as time efficient as possible.
  \item Any analysis should be able to cope with black-boxed parts of the system.
\end{enumerate}

\subsection{Graph Databases}
\label{sec:bg_graphdbs}
\subsubsection{Comparison and Selection of a Graph Database}
\label{ssec:bg_comparisongraphdbs}

The following definition of Graph Databases is provided by \cite[p.5]{graphdatabases}: "A graph database management system (henceforth, a graph database) is an online database management system with Create, Read, Update, and Delete (CRUD) methods that expose a graph data model".
They further specify graph data models as the underlying concept of how the relations between entities in the graph database are stored.
According to \cite{graphdatabases}, the property graph, the hypergraphs and Resource Description Framework Triples (RDF) are the dominant graph data models.

\cite{fernandes2018graph} compared 5 different graph database implementations, which also cover the various graph data models, which are summarized in \autoref{tab:comparisonofdatabases}.

\begin{table}[h]
  \caption{Summary of the database comparison conducted by \cite{fernandes2018graph} and supplemented by information from \cite{allegroGraphwebsite2021,Arangodbwebsite2021,infinitegraphwebsite2021}. }
  \label{tab:comparisonofdatabases}
  \begin{tabular}{|l|l|l|l|}
    \hline
    \textbf{Database} & \textbf{Graph Data Model} & \textbf{Query Language} & \textbf{Open Source} \\\hline
    AllegroGraph & RDF & SPARQL, Prolog & no\\\hline
    ArangoDB & multi-model  &  ArangoDB Query Language & community edition \\\hline
    InfiniteGraph &  Property Graph Model & "DO"  & no \\\hline
    Neo4j & Property Graph Model & Cypher & community edition \\\hline
    OrientDB &  multi-model & Gremlin, SQL  & yes \\\hline
  \end{tabular}
\end{table}

As \autoref{tab:comparisonofdatabases} shows, graph databases are not yet in a state of development where one query language emerged as prevalent over the others.

The databases in closer consideration for this work were OrientDB, ArangoDB and Neo4j, due to being the only open source databases.
Additional requirements for the choice were
\begin{itemize}
  \item available support and examples,
  \item available documentation and guidebooks,
  \item available drivers for python,
  \item ease of use and ease of installation on Windows, MacOS and Linux,
  \item capable graphical viewing tools,
  \item ability to handle up to a few million elements efficiently,
\end{itemize}

Regarding the efficiency, all of the above databases are up to the task.
Examples and documentation are also sufficiently available for all of the above databases.
As for guidebooks, Neo4j stands out with a variety of books made available for free on their website, that help beginners to get started with the database (see \cite{neo4jfordummies,neo4jalgorithms,graph_algorithms_needham,learningneo4j,graphalgouserguide}).
This as well as the ease of use, ease of installation and large number of publicly available examples built with the database were seen as especially important, since a low effort of getting started with the analysis may open the idea to a broader audience.
Since there is also a variety of tools available to query and view Neo4j graphs, we decided for Neo4j, although we do not see any obstacles to trying out other graph databases for the same kind of analyses.

\subsubsection{The Labelled Property Graph Model}
\label{ssec:bg_labelledpropertygraphmodel}
How the database stores information and makes it available to users has influence on the specification of the graph schema.
Therefore, informational constructs may require slight adaptions of the graph schema from one graph database to the other.
Neo4j employs the so called Labelled Property Graph Model, which shall briefly be described in the following(\cite{graph_algorithms_needham}).

Like any other graph database model, the main components of the Labelled Property Graph Model are nodes and edges.
Compared to other graph data models, the Labelled Property Graph Model allows to store key-value pairs as properties directly on nodes and edges, whereas other graph data models such as rdf require a separate node for every property that shall be stored and do not allow for properties of edges (\cite{neumann2011rdf,graph_algorithms_needham}).
Edges in the Labelled Property Graph Model carry a single type, describing the relation formed between the nodes.
If multiple types shall be assigned to a connection between two nodes, the nodes are connected with multiple edges.
The edges are directed, i.e. every edge has a dedicated source node and a dedicated target node.
An edge can also only ever connect exactly two nodes.
If more than two nodes shall be connected, a so-called hypernode can be employed, to which all nodes are connected that share the relation (\cite{Angles2018intrographdm}).
Apart from key-value based properties, nodes can carry multiple labels, which define the type of a node.
The application we have in mind is to investigate SysML Models with the help of graph queries.
The necessary semantics can be taken from the System Modelling Language and the specific model under investigation.
While the semantics could be styled more elaborately in  other graph models such as rdf, the gentle learning curve established by a simpler graph schema allows for easier understanding and thereby facilitates the application of the schema.

\subsection{Introduction to the Show Case Application}
\begin{figure}[!b]
  \centering
    \includegraphics[width=.5\textwidth]{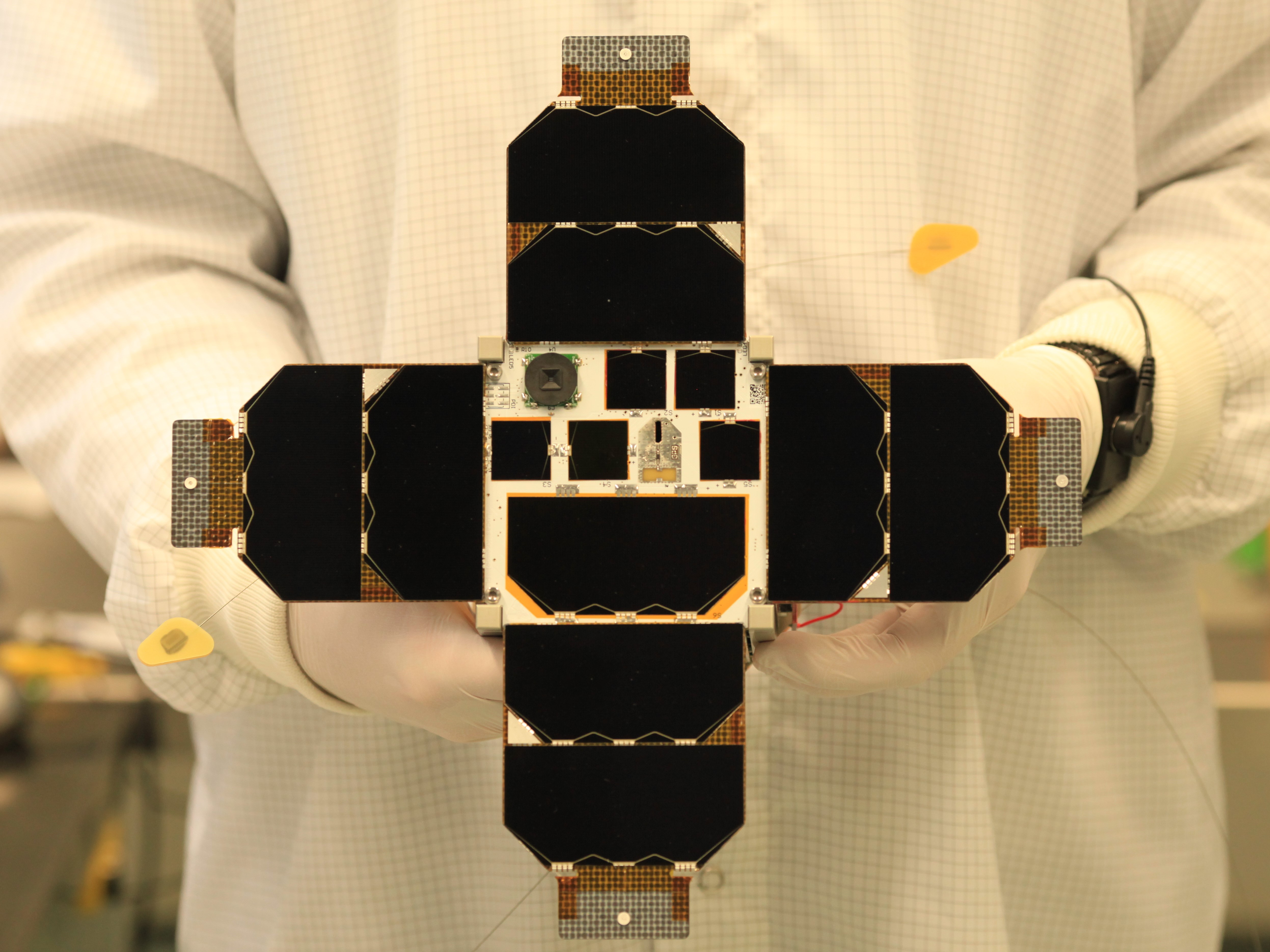}
    \caption{Photo of the MOVE-II Spacecraft showing the solar array with the payload solar cells in the middle}
    \label{fig_moveii}
\end{figure}

We use the Munich Orbital Verification Experiment II (MOVE-II) as showcase application.
The MOVE-II spacecraft is a 10x10x10cm large satellite following the CubeSat form factor (see \cite{CDS}).
The mission was founded in 2015 with the goal of hands-on practice for students looking for a career in the aerospace sector (\cite{langer2015move}).
Its payload consists of novel quatro-junction solar cell prototypes, whose degradation due to space radiation shall be measured over time (\cite{rutzinger2016solarcellpaper}).
\Cref{fig_moveii} shows the spacecraft in its deployed configuration, including the solar cell payload in the middle of the spacecraft and the four solar generators surrounding it.

After a successful launch in December 2018, the spacecraft is now in orbit since over 2.5 years and performing well (\cite{ruckerl2019first,roberts2019behind}).
The mission is a typical case of small spacecraft developments;
parts of the system such as the Command and Data Handling software, the payload, the structure and mechanisms or the Attitude Determination and Control System were designed and built in house by students, while other subsystems, such as the hardware for the Command and Data Handling system or the Electrical Power System were acquired by external vendors (\cite{langer2017move}).
The SysML model employed for the analyses in this paper covers the structure of the whole spacecraft, including any data flows and electrical flows, any sensor measurements and the satellite's software.
It furthermore covers the structure of the mission's ground segment, i.e. everything necessary to operate the satellite, such as the ground station with radio equipment and antennas or the servers and software of the operations interface and telemetry database.

\FloatBarrier
\section{Development of the Graph Schema}
\label{sec:Development of the Graph Schema}

This section provides an overview of the development of the graph schema and modelling guidelines.
In order to analyze a system by combining SysML and graph analysis, a graph schema has to be defined, with the help of which the SysML Model can be transferred from its editing tool into a labelled property graph.
A graph schema explains how nodes in a graph can be labelled and which relation types can exist between specific nodes.

\subsection{Existing Graph Schemata for SysML}
\label{sec:Existing Graph Schemata for SysML}

Two schemata are mentioned in the literature, which - before progressing to the questions set on SysML Models for graph analysis - shall be discussed briefly.

\cite{petnga2019} proposes a schema focusing on requirements analysis.
Goal of the schema is to assess completeness, consistency and correctness of requirements in a SysML model built in MagicDraw.
The author provides a detailed schema for the requirements related elements blocks, test cases and requirements.
The model falls short of considering any other element of the System Modelling Language.
Also the questions on which the schema builds are kept quite simplistic and do not exploit the strengths of a graph database.
The analyses are centered on:
\begin{itemize}
  \item What percentage of requirements are not completely defined?
  \item What percentage of requirements are not satisfied or not verified?
  \item How many elements with duplicate names exist?
\end{itemize}
None of these questions requires a graph traversal of more than one relation, i.e. they could be answered just as well by reading the elements into a table-based database or by employing table methods provided by MagicDraw.
\cite{petnga2019} further brings up the idea of applying graph algorithms to the model, which would allow finding critical elements in the system, or elements that have the largest influence on others and consequently applies the betweenness centrality algorithm on the requirements of the system, to find out which requirements have the largest influence on others.
While this yields information for the first time that could not be obtained as easily by other means, he falls short of defining what routes and elements should be selected for the algorithm, which makes a major difference in the results.

The second graph schema found in the literature is maintained by the company Intercax as part of their Syndeia software suite (see \cite{syndeiawebsite}).
First proposed in \cite{bajaj2011slim}, the idea behind Syndeia is to generate a "Total System Model", which they specify as a model that allows to link information across various information sources in a graph database.
The MBSE model is one of the information sources integrated into the total system model, alongside product lifecycle management, Computer Aided Design systems, Databases such as MySQL or Neo4j, simulation tools or application lifecycle management systems such as Jira or git (\cite{syndeiawebsite}).
The software is commercially distributed and was for example applied on the Large Synoptic Survey Telescope for Verification and Validation purposes (\cite{selvy2018v}).
Multiple publications were made on the software, its applications and involved challenges (see \cite{bajaj2016mbse++,bajaj2017graph,fisher20143}).
However, no detailed specification of the graph model itself or data of the use case systems is provided in any of the publications, making it difficult to reproduce any of the results described in the publications.
In \cite{bajaj2017graph} they show detailed graph query results, but fall short of providing the schema that allows to query the system.
While the lack of a detailed specification may be attributed to the commercial distribution of the software, a second lack in the publications on Syndeia is a SysML modelling strategy to go with the graph schema.
Analogous to \cite{petnga2019} the queries presented by \cite{bajaj2017graph} on the SysML graph are of a simplistic nature and do not exploit the potential of the graph database.
The queries range from \textit{which elements are connected to a specific element} to \textit{Which behaviors are attributed to a specific model element} and \textit{which paths exist between a specific pair of elements}.
The last query is the only one in the paper that exploits the potential of a graph database compared to relational databases.
Combining a graph schema with an appropriate modelling strategy allows for far deeper analyses and a higher quality of the information drawn from the model as is shown in the following.

\subsection{Analyses for a SysML Graph Schema}
\label{Analyses for a SysML Graph Schema}

According to \cite{van2014learning} the aim of a graph schema should be to enable answering all questions that can be foreseen to be put to the graph with a minimum of required syntax.
Therefore, this section defines questions on specific aspects of SysML models, such as structural or behavioral information.
For the sake of brevity, requirements and use cases are not addressed in this publication.
The questions provided in the following are the result of five years of personal experience in systems engineering for small spacecraft and testing and operations of small spacecraft.
The list is of course not exhaustive, but sufficient to help drafting the graph schema and modelling guidelines.

\subsubsection{Analyzing the Structural Part of a SysML Model}
\label{Analyzing the Structural Part of a SysML Model}
SysML diagrams can be separated into three groups; structural diagrams, behavioral diagrams and requirements modelling.
The following questions can be set to the structural part (i.e. Block Definition Diagrams, Internal Block Diagrams and Parameter Diagrams) of a SysML Model:
\begin{enumerate}
	\item What is component X composed of?
  \label{question_What_is_component_X_composed_of?}
	\item What types of ports are used over a certain range of equipment?
  \label{question_What_types_of_data_ports_are_used_over_a_certain_range_of_equipment?}
	\item What elements belong to a certain class?
  \label{question_What_elements_belong_to_a_certain_class?}
	\item What systems employ a certain type of component?
  \label{question_What_systems_employ_a_certain_type_of_component?}
  \item What component supplies system X with power?
  \label{question_What_component_supplies_system_X_with_power?}
  \item How is information Y processed within a certain subsystem?
  \label{question_How_is_information_Y_being_processed_within_a_certain_subsystem?}
  \item How is information Y processed globally?
  \label{question_global_datapath}
  \item Which components draw power from a certain supply component?
  \label{question_Which_components_draw_power_from_a_certain_supply_component?}
\end{enumerate}
The following questions show the might of a graph analysis, as a complete fallout analysis can be performed with the same type of queries:
\begin{enumerate}
	\setcounter{enumi}{8}
  \item What is the source of telemetry Y and what could influence its measurement?
  \label{question_What_is_the_source_of_telemetry_Y_and_what_could_influence_its_measurement?}
  \item Given an anomaly on a specific subset of a system's telemetry,  which components are most likely to have caused it? Which components can be ruled out?
  \label{question_Given_an_anomaly_on_a_specific_subset_of_a_system's_telemetry,__which_components_are_most_likely_to_have_caused_it?_Which_components_can_be_ruled_out?}
  \item Given a failure of component X, are there any alternative ways of acquiring data usually processed by component X?
  \label{question_Given_a_failure_of_component_X,_are_there_any_alternative_ways_of_acquiring_data_usually_processed_by_component_X?}
	\item What happens if component X breaks?
  \label{question_What_happens_if_component_X_breaks?}
	\begin{enumerate}
		\item Which systems will not work nominally anymore as they process data coming from component X?
    \label{question_Which systems will not work nominally anymore as they process data coming from component X?}
		\item Which components will be offline in case of an electrical short in component X?
    \label{question_short_offline_systems}
		\item Which components will suffer from a loss of input, as they depend on data processed by any of the components offline due to the electrical short in component X? \label{question_short_systems_without_proper_data_input}
	\end{enumerate}
\end{enumerate}

\subsubsection{Analyzing the Behavioral Part of a SysML Model}
\label{Analyzing the Behavioral Part of a SysML Model}
Behavioral Diagrams such as state machines, activity diagrams and sequence diagrams describe the response of a system to certain events and conditions.
They are used to model expected behavior, required inputs and task sequences to reach certain states of the system or to produce a certain output, compare \cite{friedenthal2011practical}. Analogous to this are the questions that can be answered by analyzing behavioral diagrams:
\begin{enumerate}
  \setcounter{enumi}{12}
  \item Which conditions lead to state A?
	\item What is the shortest path from state A to state B? \label{question_shortest_path_states}
	\begin{enumerate}
		  \item While condition C cannot be met? \label{question_stm_reroute_condition_cannot_be_met}
	    \item Without changing the state of equipment D? \label{question_wo_changing_equipment_D_state}
      \item Which conditions have to be fulfilled for this path? \label{question_conditions_along_stm_path}
      \item Which activities are performed along this path? \label{question_activities_along_stm_path}
	\end{enumerate}

  \item Which functions/activities require object E? \label{question_activity_object}
  \item Starting at activity F, is there a way to reach activity G? Which activities G are on that route and which alternatives could be taken? \label{question_activity_routes}
  \item Which is the shortest route from activity F to activity G? \label{question_activity_shortest_route}
  \item Which conditions have to be met on the route from activity F to G and which inputs provided? \label{question_activity_conditions_and_inputs}
  \item Which activities lead to the production of object E? \label{question_activity_activities_for_production}
  \item Which inputs are required to produce object E? \label{question_activity_inputs_for_production}
	\item In case condition C cannot be met, which states of the system cannot be reached anymore? \label{question_condition_k_cannot_be_met_which_states_cannot_be_reached_anymore}
	\item In case condition C cannot be met, which outputs cannot be acquired anymore?
  \label{question_In_case_condition_C_cannot_be_met_which_outputs_cannot_be_acquired_anymore}
\end{enumerate}

\subsection{Proposed Graph Schema}
\label{Proposed Graph Schema}
The proposed graph schema is - as the subsection before - structured according to the main aspects of SysML;
structure and behavior.
The design goal of the schema is to enable answering the questions defined in \Cref{Analyses for a SysML Graph Schema} by using Cypher queries performed on a Neo4j database.

Each paragraph describes the definition of node-labels and relation-types as well as properties stored on relations and nodes.
The labels were chosen with a focus on readability.
The idea behind this is to maintain a shallow learning curve and keep to Cypher's declarative nature.
Representations that read like a sentence are easier to understand and remember.
For example, \texttt{(Mobile Charger) -[:CLASS]-> (Power Converter)} does not make it apparent yet which is the class and which the element.
\texttt{(Mobile Charger) -[:IS\_OF\_TYPE]-> (Power Converter)} makes it clear that the Mobile Charger is an element of the class Power Converter.

\subsubsection{Graph Schema for Structural Aspects of a SysML Model}
\label{Graph Schema for Structural Aspects of a SysML Model}
Since the SysML is a graphic modelling language, most concepts can be transferred straight-forwardly (compare \cite{sysml-16}):
\begin{itemize}
  \item SysML Blocks become nodes with the label \texttt{:BLOCK}.
  \item SysML Ports become nodes with the label \texttt{:PORT}.
  \item Instances of Blocks become nodes with the labels \texttt{:BLOCK} and \texttt{:INSTANCE}
  \item Generalizations become \texttt{:IS\_OF\_TYPE} relations.
  \item Aggregations and Compositions, i.e. the associations that structure SysML Blocks hierarchically, become  \texttt{:IS\_PART\_OF} relations.
\end{itemize}

One of the ideas behind the graph schema for SysML is to avoid the need to know every aspect of the system that was ever modelled but to be able to retrieve any information via queries.
Since aggregations and compositions are both used to structure a set of blocks hierarchically, they both get the same relation type \texttt{:IS\_PART\_OF}. Details on the relation type can be retrieved via the property \texttt{type}.
The \texttt{:IS\_PART\_OF} relation is also used to connect ports to their respective blocks.

Instantiations in SysML are not always straight forward.
While every block in the diagram becomes instantiated at the moment an internal block diagram is built,  the ports connected to the blocks are not instantiated.
Furthermore, the block hosting the internal block diagram and is represented by the diagram-frame is not instantiated when an internal block diagram is built.

\begin{figure}
  \centering
  \includegraphics[width=.6\textwidth]{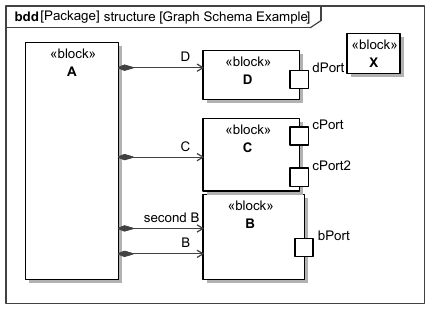}
  \caption{bdd to \Cref{fig_ibdsimpleitemflow}}
  \label{fig_bddsimpleitemflow}
\end{figure}

\begin{figure}
  \centering
  \includegraphics[width=.6\textwidth]{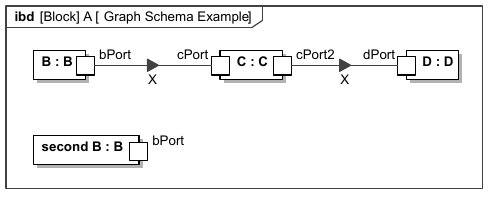}
  \caption{Example of an itemflow in an internal block diagram}
  \label{fig_ibdsimpleitemflow}
\end{figure}

This becomes relevant when looking at SysML Connectors. Connectors are used to depict connections between instances of blocks, over which physical or non-physical items can be exchanged, such as electrical power, data or physical momentum.
A concept closely related to connectors are itemflows. An itemflow describes what items are transmitted via a certain connection. The transmitted items themselves are modelled as Blocks (see \cite{sysml-16}). To distinguish them, they get the additional label \texttt{:FLOWITEM}.
\Cref{fig_ibdsimpleitemflow} shows a simple itemflow in SysML, while \Cref{fig_bddsimpleitemflow} shows the corresponding block definition diagram.
The diagram depicts four block instances, of which three are connected over respective ports.
The connectors both carry the flowitem X.
Note that the instance "second B", which is an instance of Block B does not interface any connector.

\begin{figure}
  \centering
  \includegraphics[width=.5\textwidth]{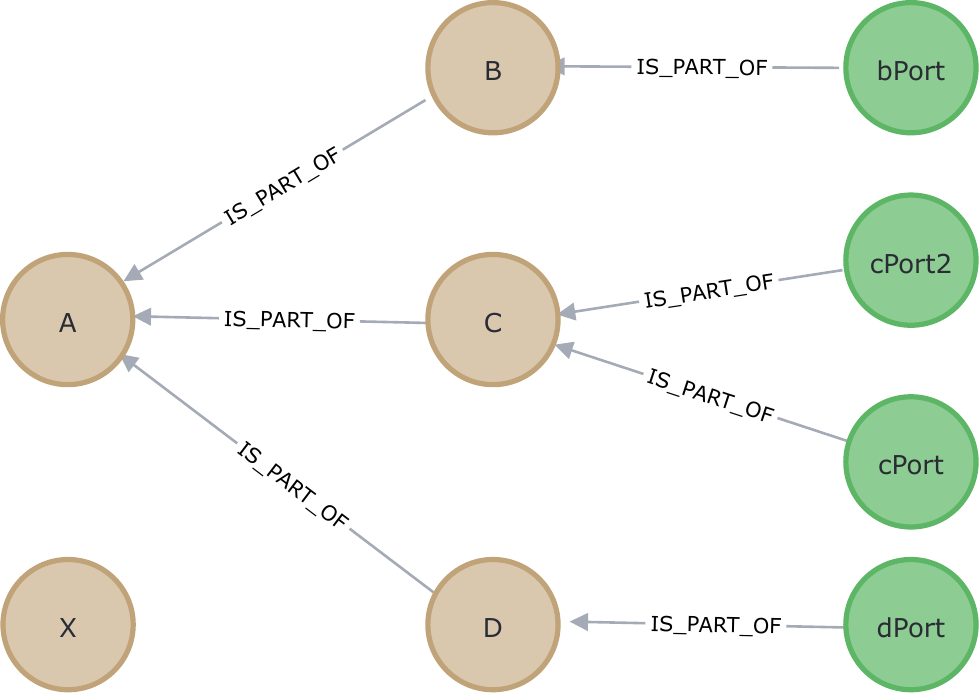}
  \caption{Graph Representation to \Cref{fig_bddsimpleitemflow}. Blocks are depicted in brown, ports in green}
  \label{fig_graphzubddsimpleitemflow}
\end{figure}

\begin{figure}
  \centering
  \includegraphics[width=.5\textwidth]{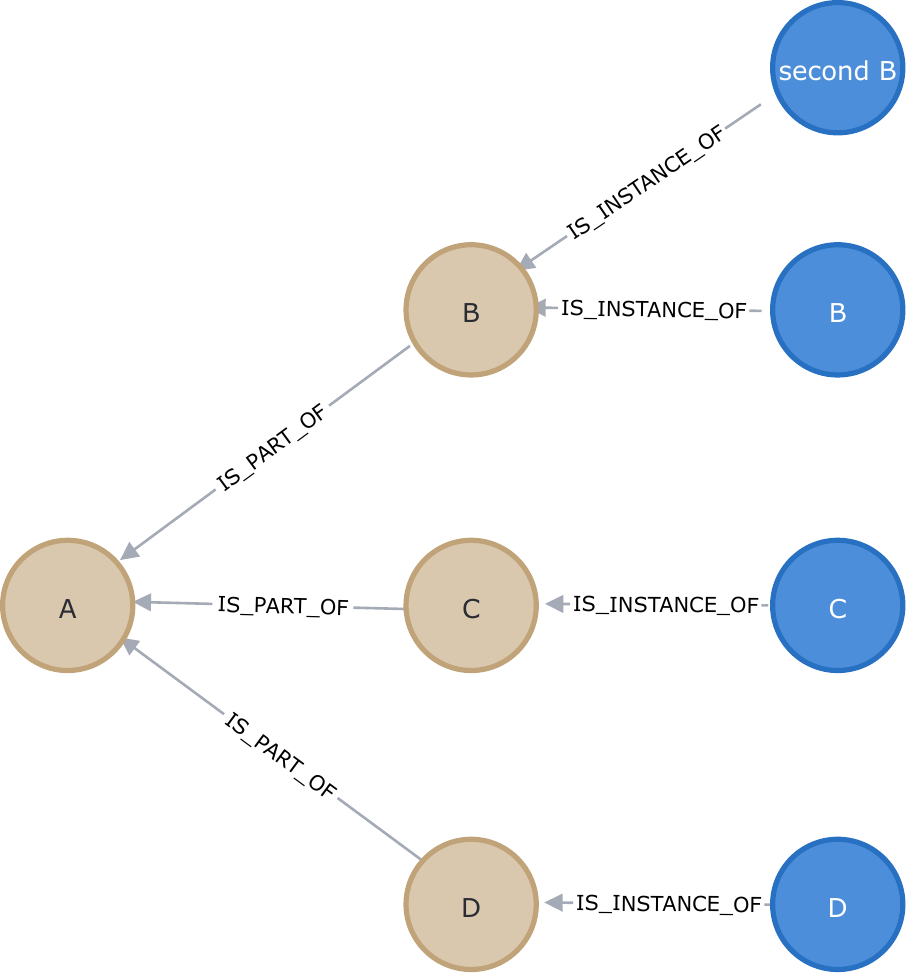}
  \caption{Graph Representation of the instances for Blocks. Blocks are depicted in brown, Instances in blue}
  \label{fig_graphzuinstanzeninsimplemitemflowibd}
\end{figure}

\begin{figure}
  \centering
  \includegraphics[width=\textwidth]{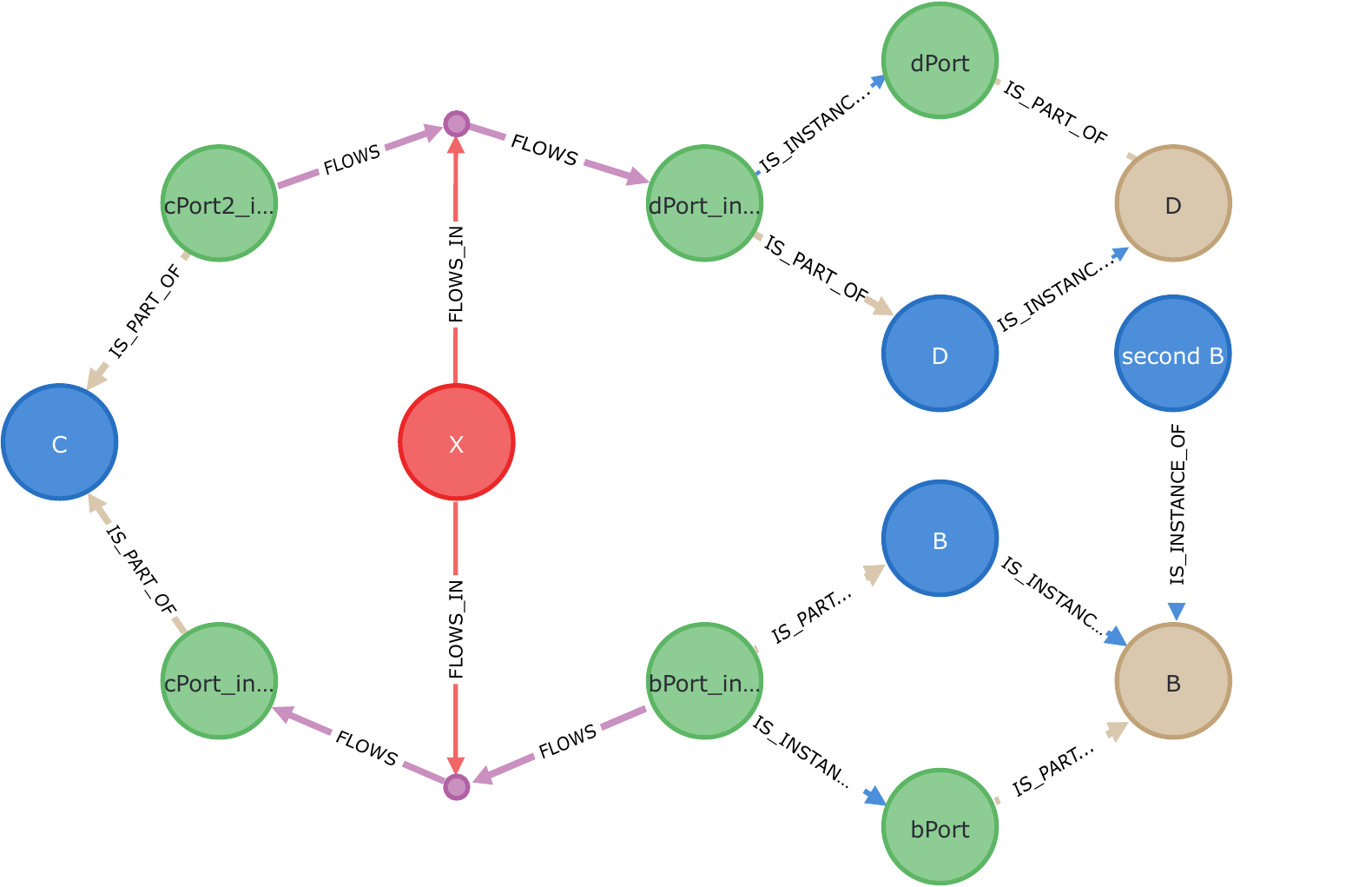}
  \caption{Graph Representation of the information related to \Cref{fig_bddsimpleitemflow} and \ref{fig_ibdsimpleitemflow}. Blocks are depicted in brown, instances of blocks in blue, hypernodes in purple, ports and port instances in green, flowitems in red. Note: As the graph itself contains all information, this is merely an excerpt showing specific information but not bound to the limits of any SysML diagram type}
  \label{fig_graphzusimplemitemflowibd}
\end{figure}

\begin{figure}
  \centering
  \includegraphics[width=.8\textwidth]{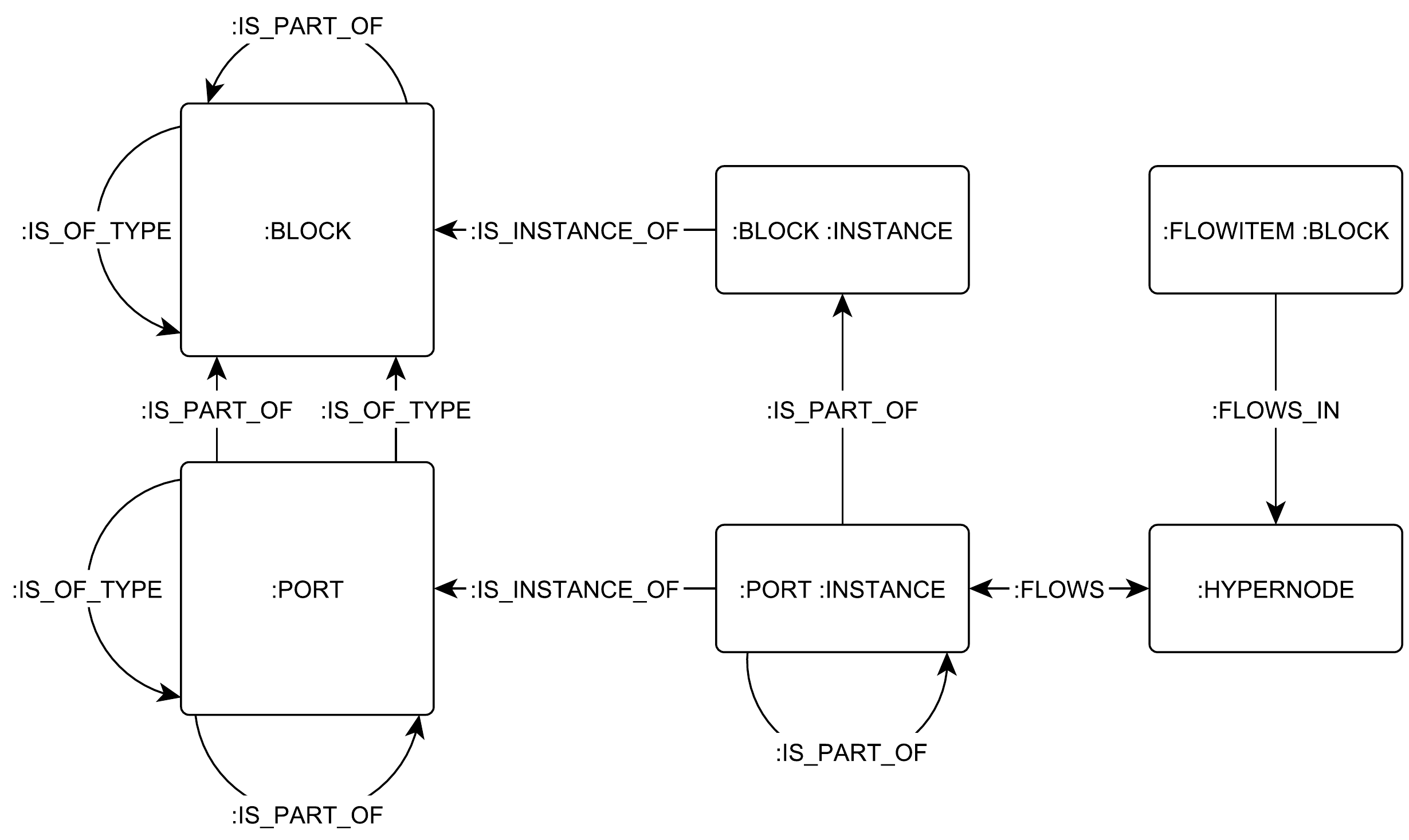}
  \caption{Proposed graph schema for Structural SysML Diagrams}
  \label{fig_str_graphschema}
\end{figure}

The graph schema has to depict the information in \Cref{fig_ibdsimpleitemflow,fig_bddsimpleitemflow} unambiguously.
\Cref{fig_graphzubddsimpleitemflow} shows the representation of the block definition diagram in \Cref{fig_bddsimpleitemflow} in the graph.
\Cref{fig_graphzuinstanzeninsimplemitemflowibd} shows the relation of instances and blocks.
\Cref{fig_graphzusimplemitemflowibd} shows the graph-transformation for the internal block diagram in \Cref{fig_ibdsimpleitemflow}.
The design is explained in the following and weighed against alternatives.

The itemflows shall be traceable through the whole system.
One possibility to translate the information into a graph is to simply create a connector from block to block and add the flowitem's name  and ID as properties on the connector.
However, this would lead to the following complications:
(a)\,the node of the flowitem has no connection to the flow as the connector runs from source block to target block and cannot be connected to a third node, (b)\,the flowitem would not be traceable anymore, if a block containing the flowitem is used to describe the itemflow  instead of the flowitem.
The next simpler solution would be to directly use the flowitem as node between the ports it flows from and to.
However, this only works until a second connector carries the same flow, as it would become untraceable which relation belongs to which connection.
The solution in \Cref{fig_graphzusimplemitemflowibd} does not have these impasses, as a connector node is created for every SysML connector. This construct is often referred to as hypernodes, which is why the node carries the label \texttt{:HYPERNODE}.

Similarly, the simplest solution to handle ports would be to omit port instantiation and only store the relation between port and block, as well as port and block-instance.
In the example shown in \Cref{fig_ibdsimpleitemflow,fig_graphzusimplemitemflowibd} however, this approach would lead to a loss of the information, whether the instance "B" or "second B" carries out the connection, as both would be connected to the port "bPort".
Hence, additionally to the elements taken from the MagicDraw model, port instances are created during the translation from MagicDraw to Neo4j to discern which port instance belongs to which block instance.
While ports can formally be instantiated in MagicDraw, requiring users to conduct this instantiation by hand and relating all port instances by hand would mean extra modelling effort that can be omitted.
\Cref{fig_str_graphschema} shows the complete graph schema for SysML Structure diagrams and to which the transformation of the internal block diagram in \Cref{fig_ibdsimpleitemflow} shown in \Cref{fig_graphzusimplemitemflowibd} is compliant.
The graphic is interpreted as follows:
Nodes are depicted in rectangulars, while relations are presented as connections between the rectangulars.
The graphic defines which combinations of labels and relations between node types are allowed by the schema.
For nodes with multiple labels (for example \texttt{:BLOCK:INSTANCE}) all relations defined by the singular labelled node (in this case \texttt{BLOCK}) are also allowed.
The relation types are written on the connections.
Arrows originating at the node they point to show relations defined between two nodes of the same type.
Additionally to the labels and relationtypes defined in \Cref{fig_str_graphschema}, all nodes carry an \texttt{id}- and a \texttt{name} property, allowing for unique identification and easy reference.
The names are taken over from the SysML model. Where no name is provided in SysML \texttt{NULL} is entered in the graph.

\subsubsection{Graph Schema for Behavioral Aspects of a SysML Model}
\label{Graph Schema for Behavioral Aspects of a SysML Model}

Behavior is modelled in SysML via Activity Diagrams, State Machines, Sequence Diagrams and Use Case Diagrams (\cite{sysml-16}).
This section will focus on Activity Diagrams, State Machines and Sequence Diagrams.

\paragraph{Graph Schema for Activity Diagrams}
\label{sec_graph_schema_act}

\begin{figure}
  \centering
  \includegraphics{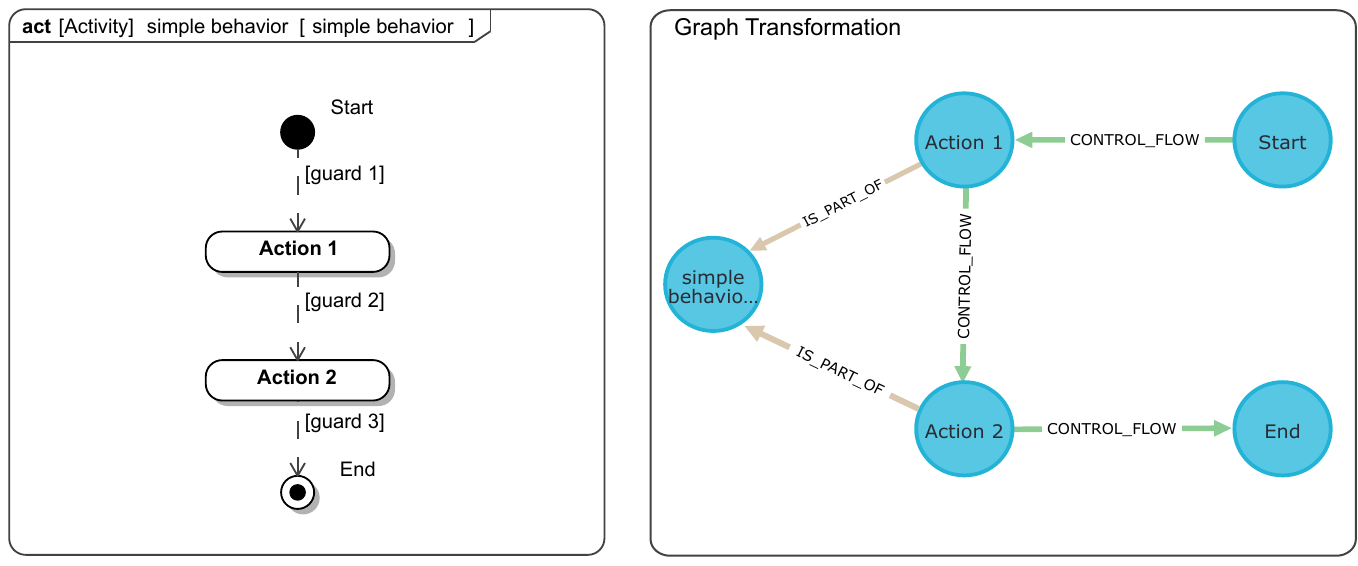}
  \caption{Simple activity diagram and its graph transformation}
  \label{fig_simple_act_and_graph_trafo}
\end{figure}

\Cref{fig_simple_act_and_graph_trafo} shows a simple activity diagram, consisting of a Start and End node and 2 activities named Action 1 and Action 2 as well as their translation into a graph.

\begin{figure}
  \centering
  \includegraphics[width=.5\textwidth]{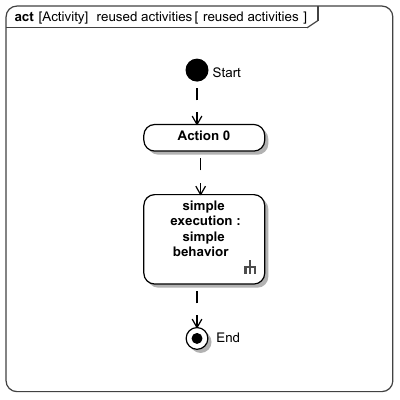}
  \caption{Nested activities}
  \label{fig_nested_act}
\end{figure}

The information contained by the control flows (dashed arrows in the SysML act diagram, green arrows in the graph transformation, see \Cref{fig_simple_act_and_graph_trafo}) is stored as properties of the \texttt{:CONTROL\_FLOW} relations, which are not depicted in \Cref{fig_simple_act_and_graph_trafo}.
Any activity diagram can be seen as an activity by itself.
Therefore, the node to the very left of the graph transformation in \Cref{fig_simple_act_and_graph_trafo} represents the diagram itself.
This becomes a necessity when dealing with nested activities.
\Cref{fig_nested_act} shows the activity \textit{simple behavior} from \Cref{fig_simple_act_and_graph_trafo} nested in another activity with the title \textit{reused activities}.
Semantically, this means \textit{simple behavior} is executed in the \textit{reused activities} diagram and the complete content of \textit{simple behavior} is run through at the time the activity is called.

\begin{figure}
  \centering
  \includegraphics{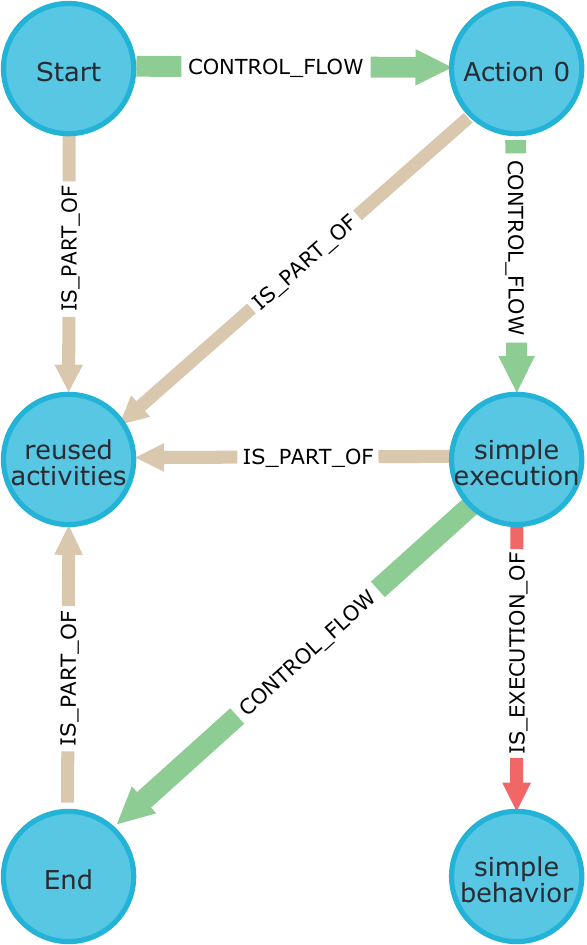}
  \caption{Nested activities transformation Alternative 1}
  \label{fig_nested_act_trafo_poss1}
\end{figure}

\begin{figure}
  \centering
  \includegraphics{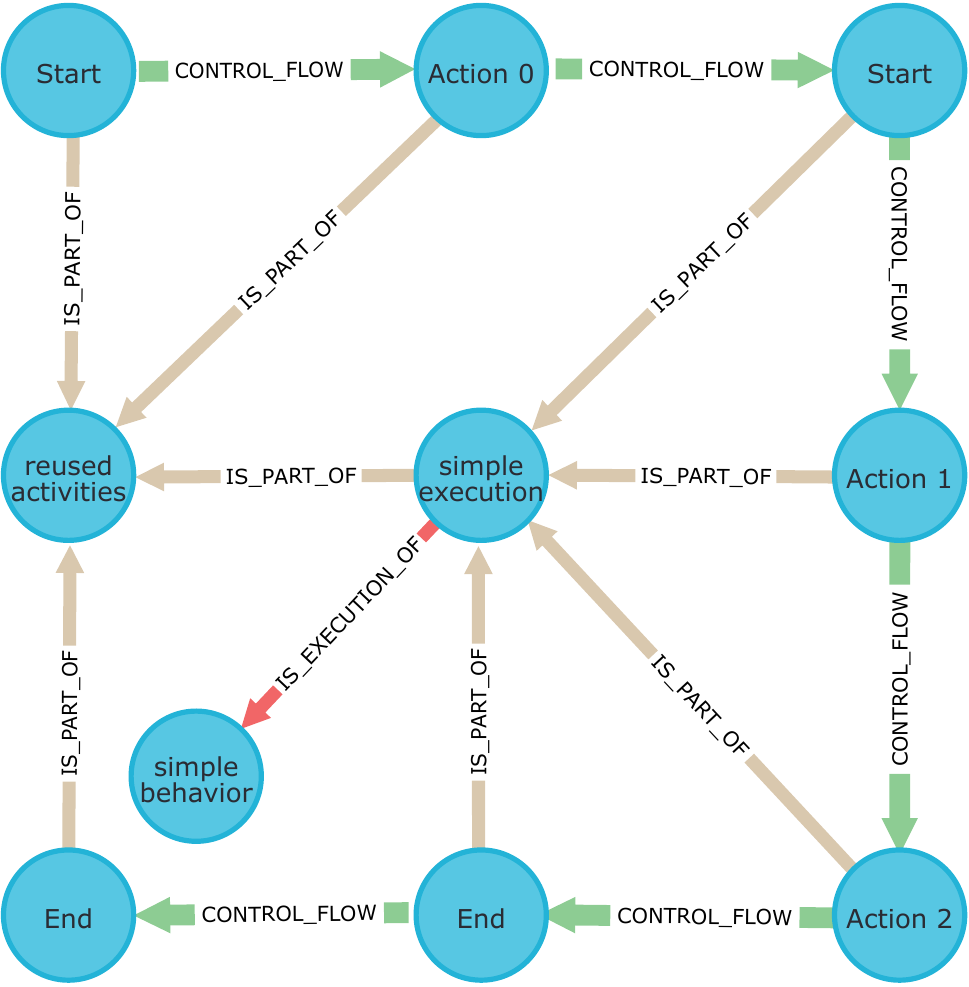}
  \caption{Nested activities transformation Alternative 2. \textit{Note:} \texttt{:IS \_INSTANCE\_OF} relations and respective nodes are omitted in this view for better readability}
  \label{fig_nested_act_trafo_poss2}
\end{figure}

\Cref{fig_nested_act_trafo_poss1,fig_nested_act_trafo_poss2} show different possibilities of the diagram's transformation.
\Cref{fig_nested_act_trafo_poss1} shows a simpler transformation that is closer to the SysML's own syntax. The \textit{simple execution} of \textit{simple behavior} is executed and connected via the corresponding control flows as depicted in \Cref{fig_nested_act}.
\Cref{fig_nested_act_trafo_poss2} shows a more complex transformation.
Each activity and node contained in the \textit{simple behavior} diagram is realized again together with the respective control flows.
The control flows that connected to \textit{simple execution} are redirected to the \textit{Start} and \textit{End} node of \textit{simple execution}.
Thereby the complete activity flow is generated and can be queried to understand, which activities have to be performed in order to achieve a certain result.

While both transformation schemata have their advantages, we decided in favor of the variation shown in \Cref{fig_nested_act_trafo_poss2}, as it is easier to understand a complete flow of activities and keeps queries simpler.

\begin{figure}
  \centering
  \includegraphics{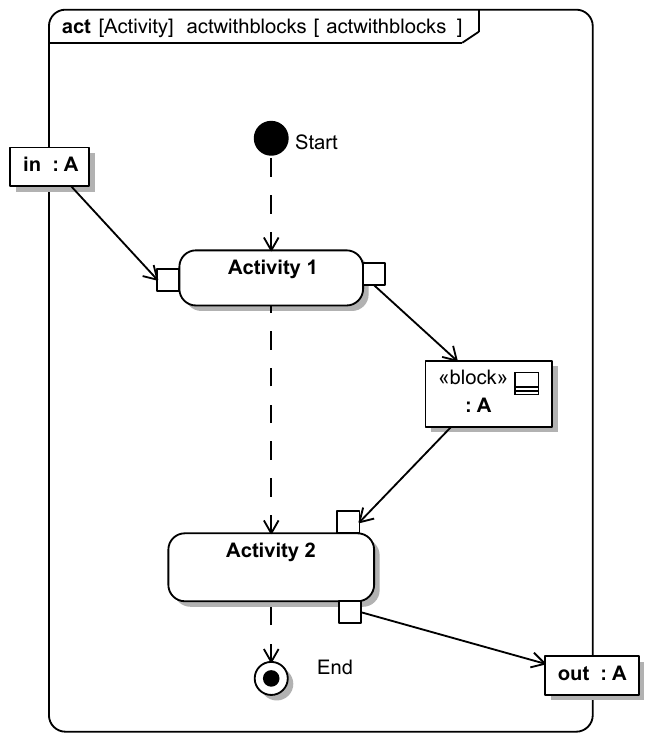}
  \caption{Activity Diagram with Blocks}
  \label{fig_act_with_blocks}
\end{figure}

Another construct of SysML Activity Diagrams is the use of blocks as inputs and outputs of activities as depicted in \Cref{fig_act_with_blocks}.
Blocks are depicted in activity diagrams as rectangles with pointed corners.
They are connected to activities in the diagram via object flows (solid arrows in \Cref{fig_act_with_blocks}).

\begin{figure}
  \centering
  \includegraphics[width=\textwidth]{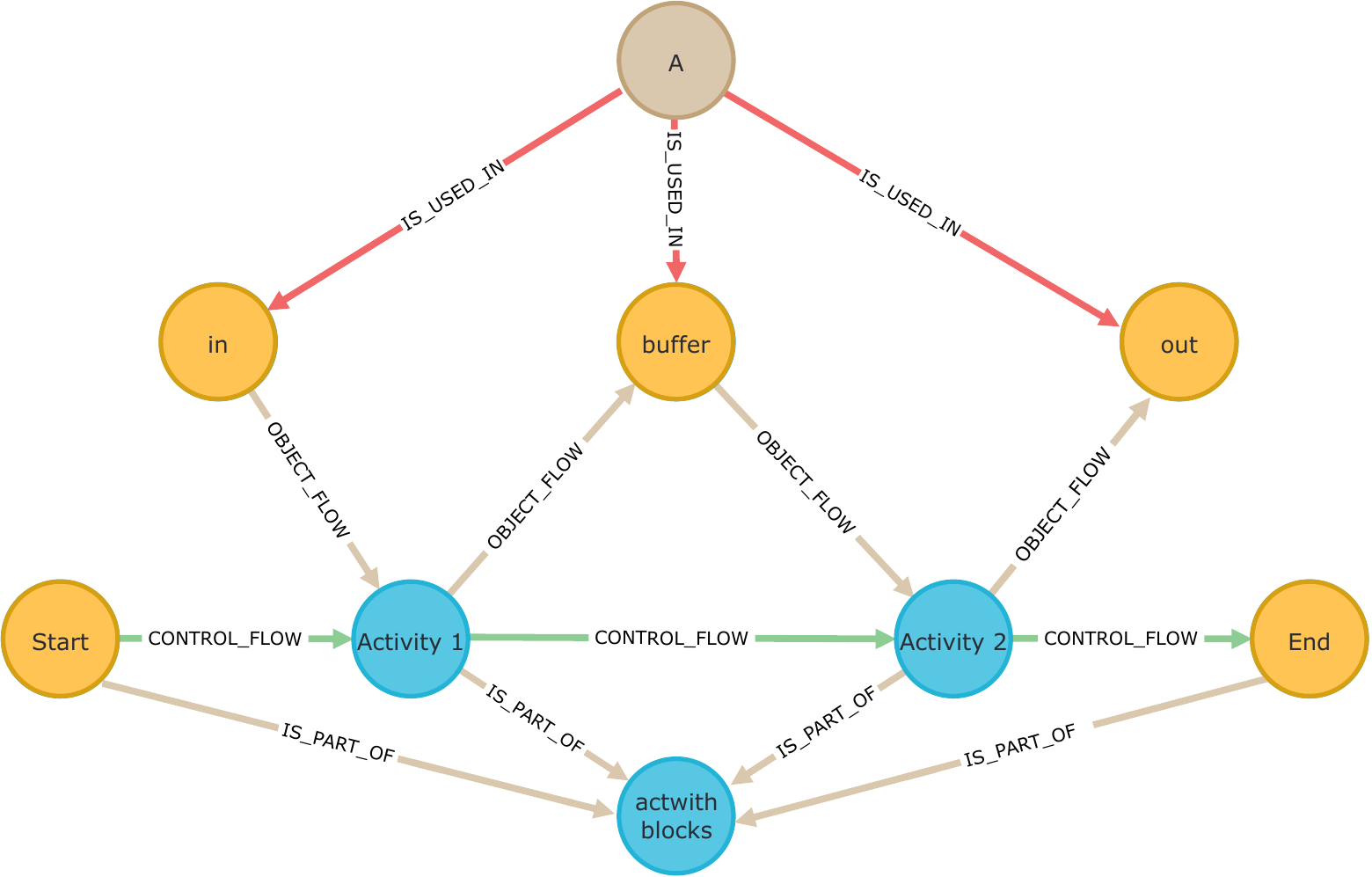}
  \caption{Graph Transformation of the Activity Diagram with Blocks in \Cref{fig_act_with_blocks}. Yellow: \texttt{:ACTNODE}, blue: \texttt{:ACTIVITY}, beige: \texttt{:BLOCK}}
  \label{fig_act_with_blocks_graph_trafo}
\end{figure}

\Cref{fig_act_with_blocks_graph_trafo} shows the diagram's translation into the graph.
To realize the usage of blocks in activity diagrams, two additional relation types are introduced: \texttt{:OBJECT\_FLOW}, which is a direct translation of its SysML counterpart, and \texttt{:IS\_USED\_IN}, which represents the connection between the respective block (\textit{A}) and the \texttt{:ACTNODE}s (\textit{in, buffer, out}) of the object flow.
The \texttt{:ACTNODE} label is also used for Initial and Final nodes, as well as Decision nodes, Fork nodes and Merge nodes.
To discern the different types, the property \texttt{nodetype} is defined, which carries the respective information.

\begin{figure}
  \centering
  \includegraphics[width=.7\textwidth]{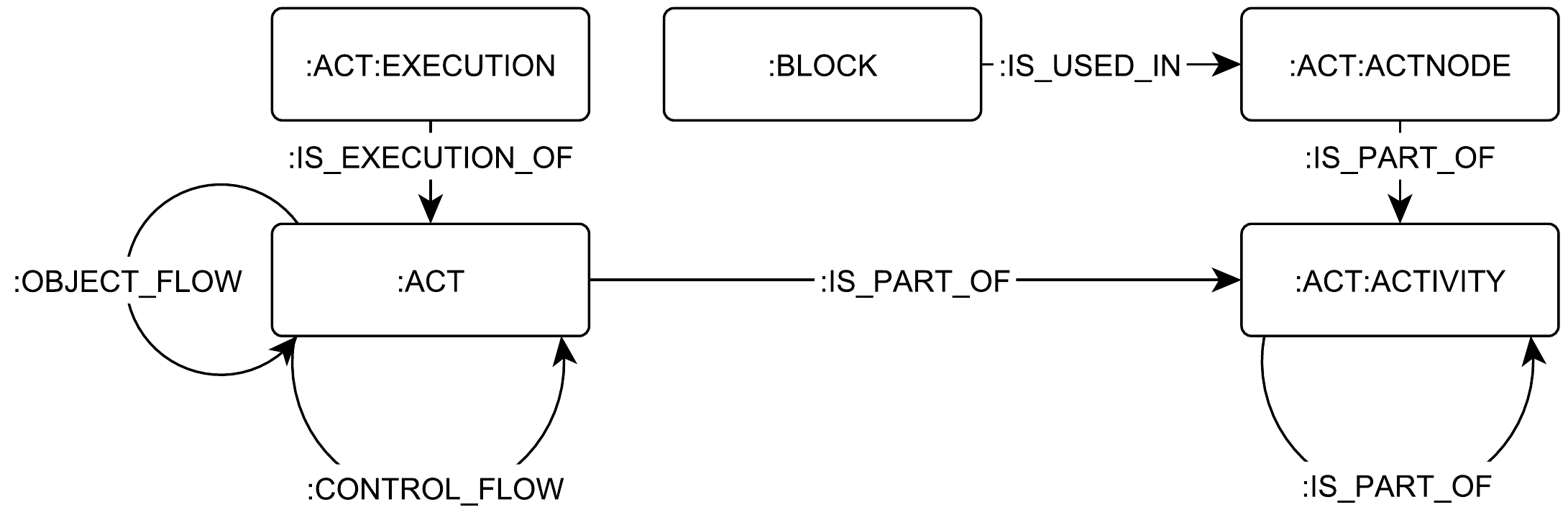}
  \caption{Graph Schema for Activity Diagrams.}
  \label{fig_act_graph_schema}
\end{figure}

This brings us to the declaration of node labels for activity graphs.
\Cref{fig_act_graph_schema} summarizes the graph schema for activity diagrams.
Discerning activities from other nodes such as Initial, End, Fork or Buffer nodes in an activity diagram is necessary to answer the Questions \ref{question_activity_routes} to \ref{question_activity_inputs_for_production} in \Cref{Analyzing the Behavioral Part of a SysML Model}.
Therefore, additionally to the \texttt{:ACTNODE} label, an \texttt{:ACTIVITY} label is defined, which is used for the actual activities.
To reduce the effort to build queries, nodes with the label \texttt{:ACTNODE} or  \texttt{:ACTIVITY} also carry the label \texttt{:ACT}, which marks any node from an Activity Diagram.
For any execution, the \texttt{:EXECUTION} label is provided, together with the relation \texttt{:IS\_EXECUTION\_OF}.
The \texttt{:BLOCK} label used in \Cref{fig_act_graph_schema} is the same one as in the graph schema for block definition diagrams and links the two diagram types.

\paragraph{Graph Schema for State Machines}

\begin{figure}[!]
  \centering
  \includegraphics{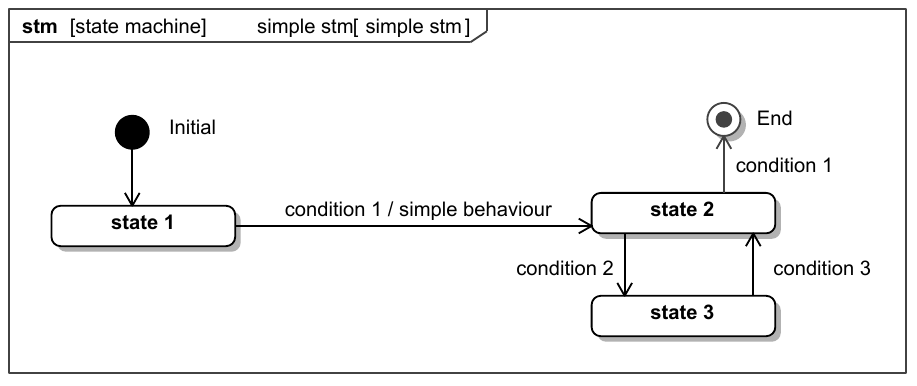}
  \caption{Simple State Machine}
  \label{fig_simple_stm}
\end{figure}

\begin{figure}[!]
  \centering
  \includegraphics{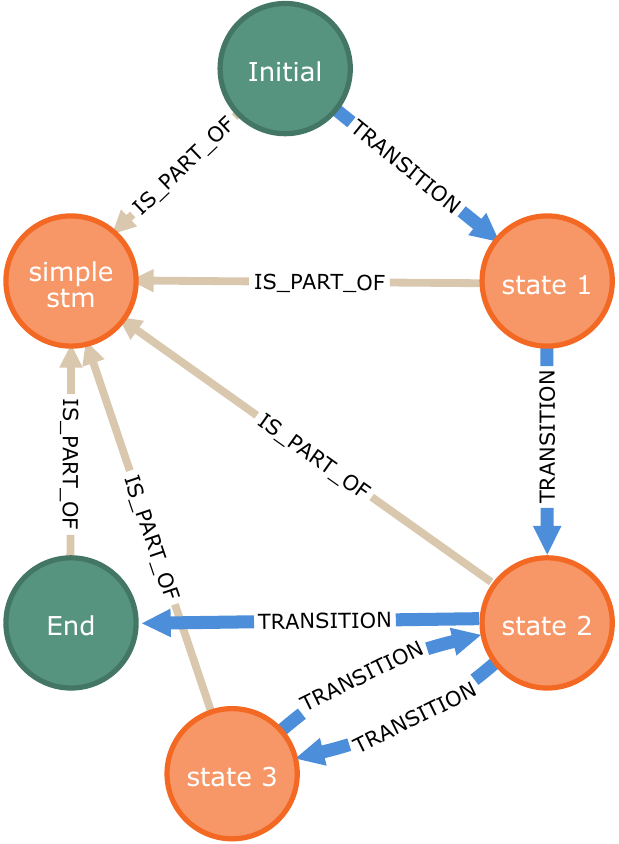}
  \caption{Possible graph transformation of the state machine in Figure \ref{fig_simple_stm}. \texttt{:STATE} nodes in orange, \texttt{:PSEUDOSTATE} nodes in green, \texttt{:TRANSITION} relations in blue, \texttt{:IS\_PART\_OF} relations in brown}
  \label{fig_simple_stm_trafo}
\end{figure}

The second important part of behavioral modelling is state machines.
\Cref{fig_simple_stm} shows a simple state machine with three states, \textit{state 1, state 2, state 3} and a set of transitions.
According to \cite[p. 162]{sysml-16}, transitions can carry triggers, guards and activities.
Triggers and guards specify when the transition becomes active, while the activity simply states that an activity shall be performed when the transition activates.
\Cref{fig_simple_stm_trafo} shows a possible transformation of the state machine in \Cref{fig_simple_stm} into a graph.
However, the schema applied in \Cref{fig_simple_stm_trafo} does not allow for complex queries on transition conditions and behavior such as Question \ref{question_shortest_path_states}
in \Cref{Analyzing the Behavioral Part of a SysML Model}.
This is resolved by a \texttt{:HYPERNODE} construct, equivalent to the \texttt{:HYPERNODE} in \Cref{Graph Schema for Structural Aspects of a SysML Model}, which is linked with the state that was left and the state which is entered via a \texttt{:TRANSITION} relation.
Triggers for state transitions, such as condition 1, 2 and 3 in \Cref{fig_simple_stm} carry the label \texttt{:TRIGGER}, which may be added additionally to any other label the conditioning node carries.
\texttt{:TRIGGERS} relations link the \texttt{:TRIGGER} with the \texttt{:HYPERNODE}.

\begin{figure}[!]
  \centering
  \includegraphics{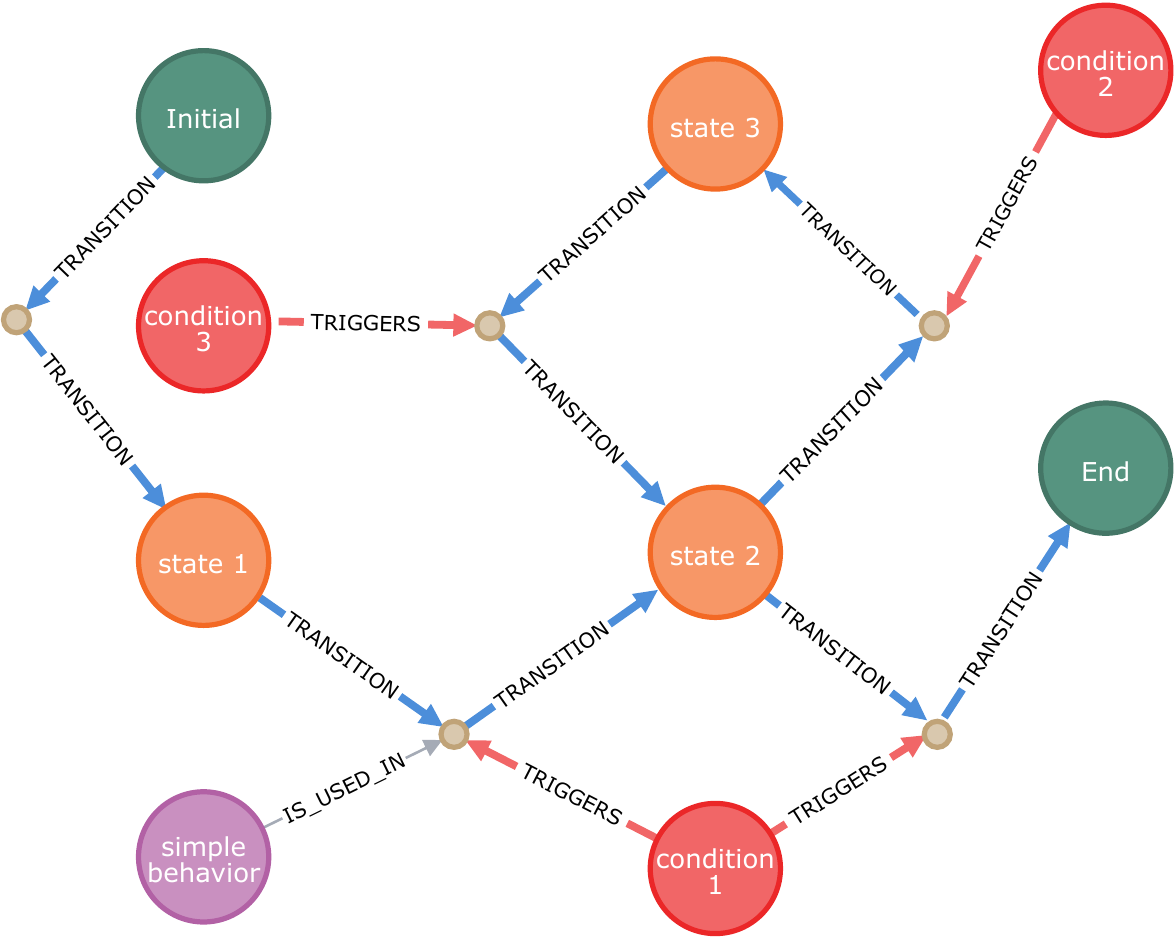}
  \caption{Graph transformation of the state machine in Figure \ref{fig_simple_stm}. \texttt{:STATE} nodes in orange, \texttt{:PSEUDOSTATE} nodes in green, \texttt{:HYPERNODE}s in brown, \texttt{:ACTIVITY} nodes in purple, \texttt{:TRIGGER} nodes in red. \textit{Note:} The node "simple stm" is not shown for better readability}
  \label{fig_simple_stm_trafo_v2}
\end{figure}

Additionally to the trigger events, so called guards can be specified to refine the conditions under which a state transition is activated.
The definition of trigger events as separate nodes is sufficient to answer the questions defined in \Cref{Analyzing the Behavioral Part of a SysML Model} though.
Therefore, guards are for now treated as properties of the \texttt{:HYPERNODE} (\texttt{guard}-property).
For an exemplary transformation of \Cref{fig_simple_stm} see \Cref{fig_simple_stm_trafo_v2}.
Note how \textit{condition 1} is used in both transitions and the activity \textit{simple behavior} is also linked to the transition, enabling us to answer questions such as Question \ref{question_shortest_path_states}\,a,\,c (What is the shortest path from state A to state B, while condition C cannot be met and which conditions have to be fulfilled for this path),
 \ref{question_condition_k_cannot_be_met_which_states_cannot_be_reached_anymore} (In case condition C cannot be met, which states of the system cannot be reached anymore) or
 \ref{question_In_case_condition_C_cannot_be_met_which_outputs_cannot_be_acquired_anymore} (In case condition C cannot be met, which outputs and signals cannot be acquired anymore).

\FloatBarrier

\begin{figure}
  \centering
  \includegraphics{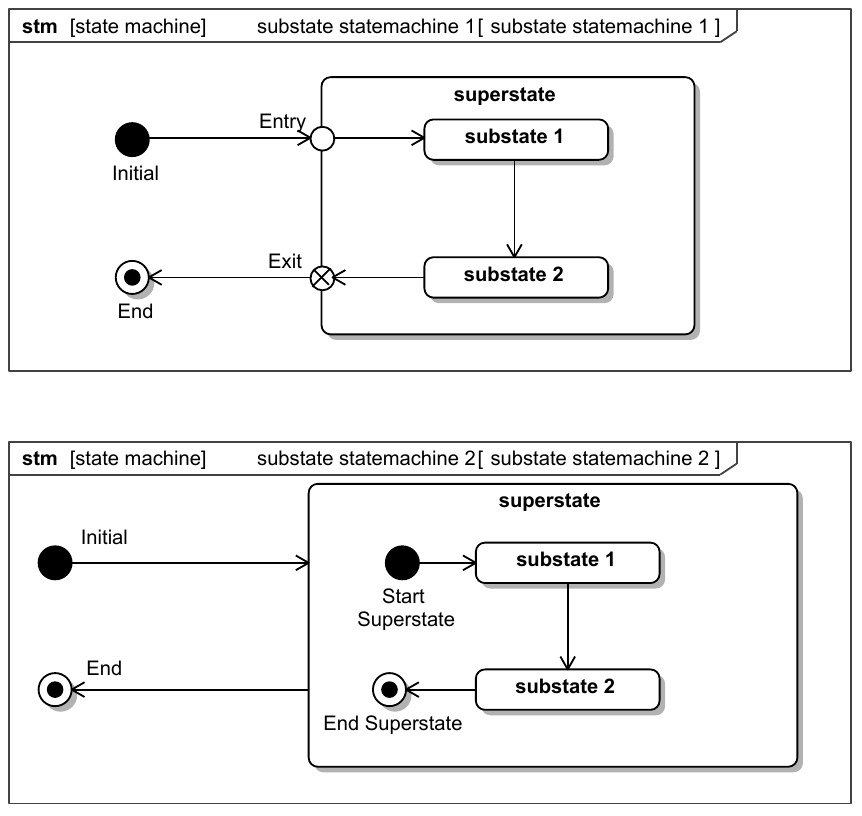}
  \caption{State machine with substates in two variations}
  \label{fig_stm_with_substates}
\end{figure}

\begin{figure}
  \centering
  \includegraphics{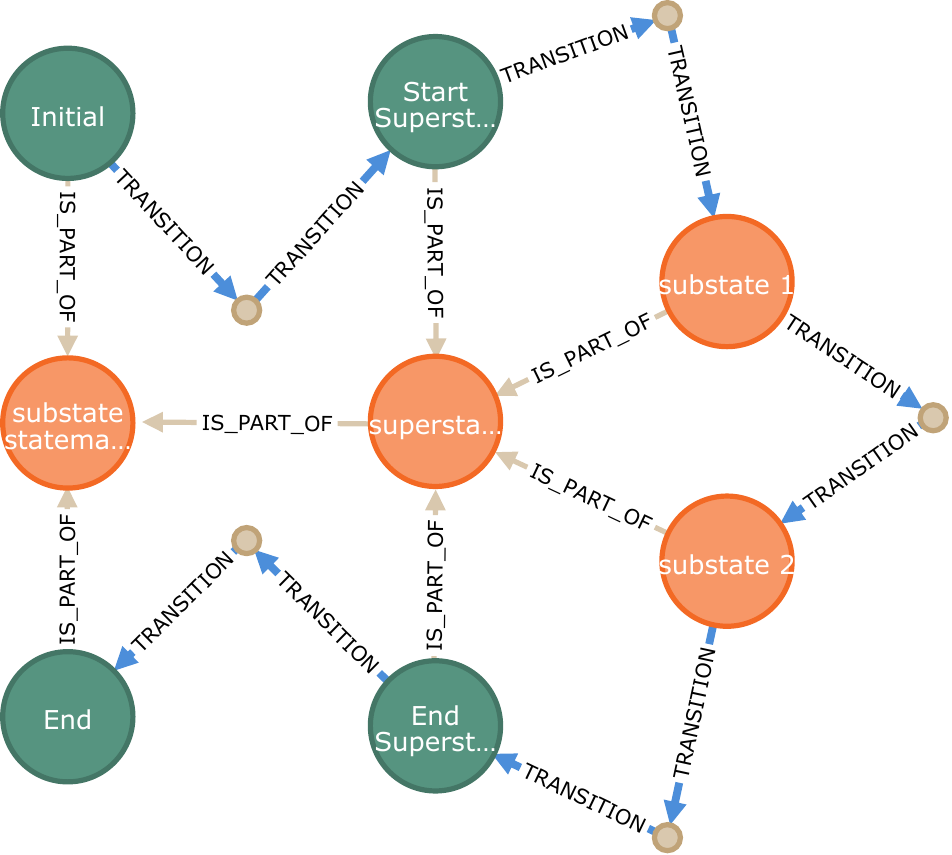}
  \caption{Graph Transformation of the state machine with substates in \Cref{fig_stm_with_substates}. Orange nodes carry a \texttt{:STATE}-label, green nodes a \texttt{:PSEUDOSTATE}-label and beige nodes a \texttt{:HYPERNODE}-label. State transitions are shown as blue relations, whereas \texttt{:IS\_PART\_OF} relations are shown in beige}
  \label{fig_stm_with_substates_graph_trafo}
\end{figure}

\begin{figure}
  \centering
  \includegraphics[width = \textwidth]{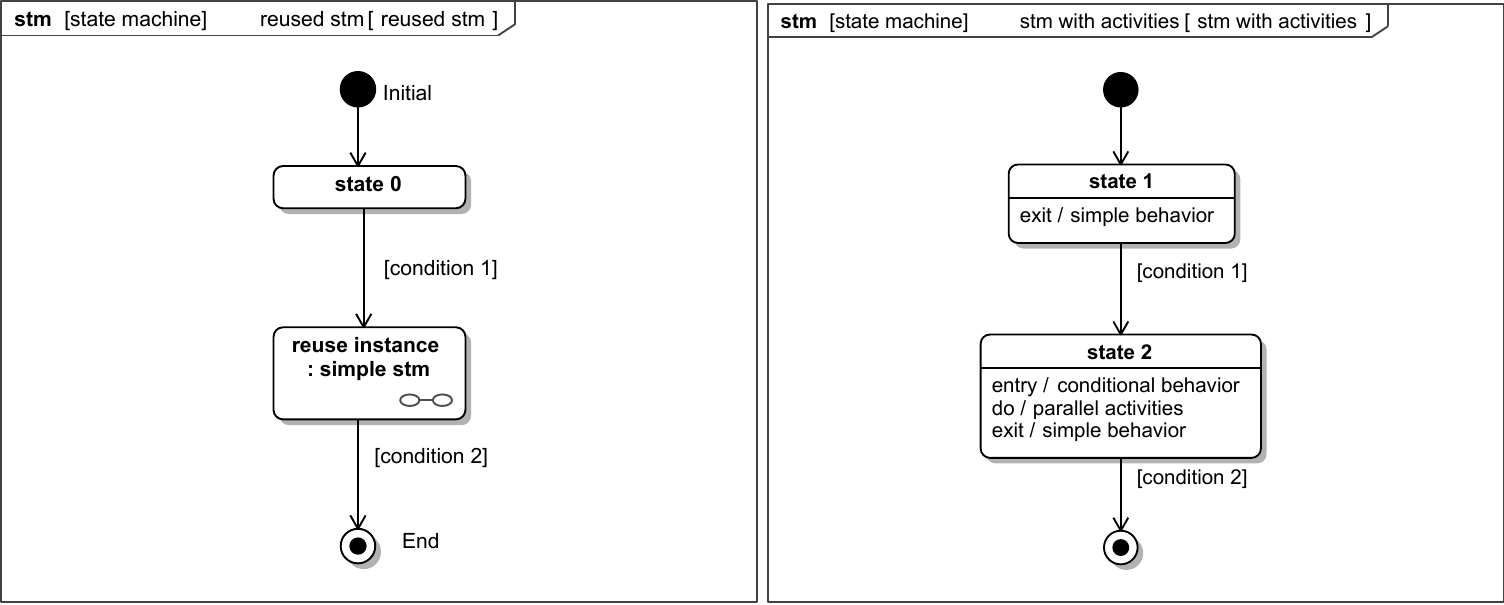}
  \caption{Left: State machine with reference to another stm. Right: State machine with activities}
  \label{fig_stm_reused}
\end{figure}

\Cref{fig_stm_with_substates} and \Cref{fig_stm_reused} show more possibilities to be considered for the transformation.
\Cref{fig_stm_with_substates} depicts state machines with substates in two variations.
This opens up possibilities for the flow of transitions, as the transition entering \textit{superstate} can either be drawn to \textit{superstate}, \textit{start superstate} (or \textit{entry}), or both.
The same applies for the transitions to leave the superstate.
Arguments can be made for and against all three variations:
\begin{itemize}
  \item Drawing the transition from \textit{initial} to \textit{superstate} allows to answer the question \textit{"How can I enter or leave superstate?"}, but makes it more cumbersome to follow the flow of transitions through the project, as for every state we now need to query whether it has a set of substates and if so include these in the list of states and transitions.
  Which level of detail is appropriate may be a difficult to answer question especially since the graph queries shall be usable without prior knowledge of the complete SysML Model.
  \item Drawing the transition from \textit{Initial} to \textit{Entry} allows to follow the transitional flow through the whole diagram, without ever querying any other relation than \texttt{:TRANSITION}.  A disadvantage of this approach is that it becomes more difficult to query how to enter and leave the \textit{superstate}.
  \item Drawing both transitions leaves us without any cumbersome queries for either way of modelling, but results in a graph that is harder to understand, since the transitions in the graph would show that you are either in the \textit{superstate} or in \textit{substate 1} or \textit{substate 2.}
\end{itemize}
Following a transitions network through a set of state machines is a more complex task than answering the question of possible entries and exits to a single state.
Therefore, I decided to stick with the second option from above.
\Cref{fig_stm_with_substates_graph_trafo} shows the graph transformation of the state machines in \Cref{fig_stm_with_substates}.

An interesting feature of state machines in SysML is their connection to activities.
Activities can be set to be performed on transitions, entry, do [while] or  exit of a state \cite[p. 161]{sysml-16}.
Such activities may for example log the information "state was left", or saving the results at the end of a measurement accumulation.
Such a state machine is displayed in \Cref{fig_stm_reused} on the right.
Employing activities is a basic feature of state machines and depicting this kind of crosscutting connection from from one diagram type to the other is one of the strengths of graph databases.
\textit{State 2} in \Cref{fig_stm_reused} shows activities  at various positions within a state (entry, do, exit).
The activities used in the state machine have to be defined elsewhere and are then only used in the state.
Using a set of activities within a state implicitly creates a new set of control flows between the entry, do and exit activities.
The graph schema covers this implicit connection by explicitly linking the three elements with \texttt{:CONTROL\_FLOW} construct (see \Cref{sec_graph_schema_act}).
Another relation is necessary between the state and the activities performed within it.
This is covered by a \texttt{:IS\_PART\_OF} relation.
While it could be argued, that a new relation type instead of reusing a relation from the structural aspects of SysML makes sense, goal of the schema is to keep it as simple as possible.
Creating new relation types for the same kind of information as in the structural part of the SysML is not necessary as the context provided by the node labels (\texttt{:BLOCK} in the structural part of the schema, \texttt{:ACTIVITY} and \texttt{:STATE} in the  behavioral part) resolves any ambiguity.

The diagram on the left side of \Cref{fig_stm_reused} shows the reused state machine \textit{reused execution: simple stm}.
The concept of nesting and reusing defined behavioral elements was already described in the beginning of this section for activities.
The same logic applies here.
Porting the execution of the stm that was already created by MagicDraw over into the graph allows for  clear and understandable semantics. Therefore, executions of states are referred to with the label \texttt{:EXECUTION} additionally to the \texttt{:STATE} label.
Apart from the \texttt{:STATE} nodes, other elements of the state machine also have to be transferred into the graph, such as initial nodes and end nodes, decision nodes, etc.
For these elements the \texttt{:PSEUDOSTATE} label is provided.

\begin{figure}
  \centering
  \includegraphics[width = .8\textwidth]{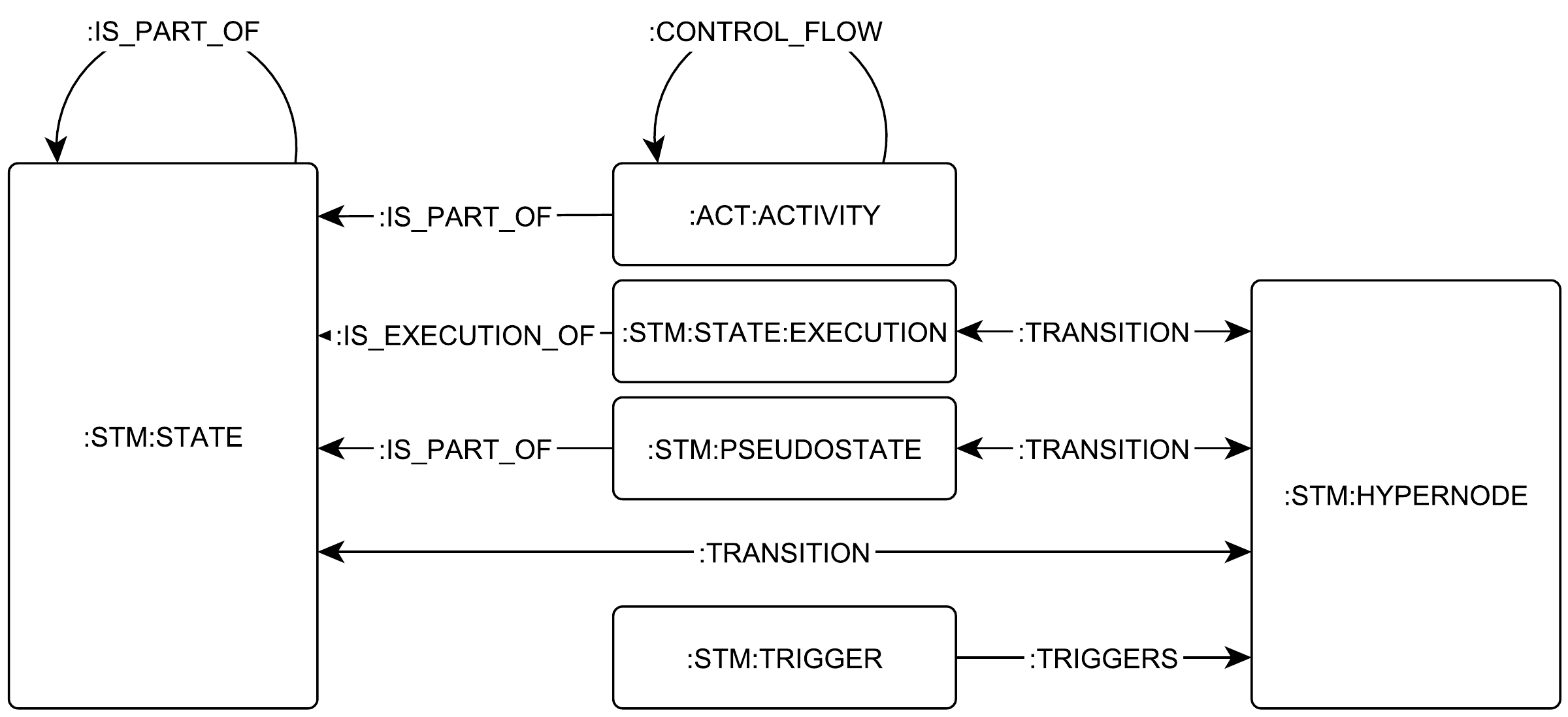}
  \caption{Graph Schema for SysML state machines}
  \label{fig_stm_graph_schema}
\end{figure}

\Cref{fig_stm_graph_schema} shows the graph schema for state machines in detail.
Nodes of the type \texttt{:STATE} and \texttt{:PSEUDOSTATE} can be connected to \texttt{:HYPERNODE}s via the \texttt{:TRANSITION} relation.
\texttt{:STATE:EXECUTION} nodes are connected to \texttt{:STATE} nodes via the \texttt{:IS\_EXECUTION\_OF} relation type.
States can also be part of other states, and therefore same as \texttt{:ACTIVITY} nodes can be connected to states via \texttt{:IS\_PART\_OF} relations.
\texttt{:TRIGGER} nodes define the conditions which are necessary to conduct a transition and relate via the \texttt{:TRIGGERS} relation to \texttt{:HYPERNODE}s.


\FloatBarrier
\section{Modelling Guidelines}
\label{sec_modelling_guideline}

One goal of the graph schema is to enable queries on a model that is not yet complete and thereby enabling graph-queries on SysML models from an early stage of development or on systems that are only partially modelled in later stages.

One goal of this set of modelling guidelines is to show how to build a SysML model that is effective for employing graph queries.
A second goal of the guidelines is to explain how to trim the guidelines, i.e. what happens if a certain rule is not followed.
When trimming these guidelines and working with incomplete models consider that \textit{the graph can only relay information that is available in the model.}
This is an important principle to understand which information should be included in the SysML model and guides us through the following section.
The last goal of this set of modelling guidelines is to be useful, which requires it to be simple and efficient to apply and to yield results fast, while touching as little as possible on other modelling principles that may be followed by the modeler.
Therefore, examples illustrating a guideline are presented at various points.

The buildup of this section follows the same principle as the previous sections, starting with structural diagrams and progressing to behavioral diagrams.

\subsection{Modelling Guidelines for Structural Aspects of a SysML-Model}
The following section refers to the questions defined in \Cref{sec:Development of the Graph Schema}. Any question number cited in the following refers to \Cref{sec:Development of the Graph Schema}.

\subsubsection{Modelling of Structural Associations}
Recalling the questions defined in \Cref{Analyzing the Structural Part of a SysML Model}, the Questions \ref{question_What_is_component_X_composed_of?} (detailing the composition of a block)  to
\ref{question_What_systems_employ_a_certain_type_of_component?} (detailing systems that employ a certain component) require a sound use of associations between blocks.
I.e. \textit{Every Block that is instantiated should be part of a structure of SharedAssociations and PartAssociations which start at the system of systems and reach any components used in the system}.
This is also required for Question \ref{question_Which_components_draw_power_from_a_certain_supply_component?} (querying components dependant on a certain power supply) or
Question \ref{question_What_happens_if_component_X_breaks?} (querying the fallout caused by a broken component).
As the graph schema unifies the concepts of part associations and shared associations into \texttt{:IS\_PART\_OF}-relations, no distinction is made in the guideline between the two association types.

The same applies for flowitems such as data: modelling flowitems as a set of associated blocks saves modelling effort when pursuing questions such as "How is information Y processed globally?"  (Question \ref{question_global_datapath}) or searching for likely culprits in case of an anomaly on a specific subset of telemetry
(Question \ref{question_Given_an_anomaly_on_a_specific_subset_of_a_system's_telemetry,__which_components_are_most_likely_to_have_caused_it?_Which_components_can_be_ruled_out?}).

\subsubsection{Generalizations}
\label{sec_generalizations_modelling_guideline}
The Questions \ref{question_What_types_of_data_ports_are_used_over_a_certain_range_of_equipment?} (querying the types of ports used over a certain range of equipment), \ref{question_What_elements_belong_to_a_certain_class?} (querying the elements belonging to a certain class) and
\ref{question_What_systems_employ_a_certain_type_of_component?} (querying which systems employ a certain type of component) are of interest when the failure of a certain component can be traced back to its working principle and according changes need to be made across the whole system.
Furthermore generalizations can be used to classify flowitems in the model into categories such as currents and voltages or data.
This distinction is necessary to answer questions regarding how a certain piece of information is processed (Questions \ref{question_How_is_information_Y_being_processed_within_a_certain_subsystem?},
\ref{question_global_datapath},
\ref{question_What_is_the_source_of_telemetry_Y_and_what_could_influence_its_measurement?},
\ref{question_Given_an_anomaly_on_a_specific_subset_of_a_system's_telemetry,__which_components_are_most_likely_to_have_caused_it?_Which_components_can_be_ruled_out?}
\ref{question_Given_a_failure_of_component_X,_are_there_any_alternative_ways_of_acquiring_data_usually_processed_by_component_X?}) or
regarding  power paths (Questions
\ref{question_Which_components_draw_power_from_a_certain_supply_component?} and
\ref{question_What_happens_if_component_X_breaks?} (a) to (c) ).

All of these questions require the use of generalizations as defined in \cite{sysml-16}.
A hierarchy in which port types, component types and data types are defined is therefore sensible, but can be limited to the level of detail that allows to answer the above questions.
On some systems this may require to refine them to the point where the protocol of the port is defined (for example CAN, USB, Ethernet), while for other systems a simple differentiation between analog and digital ports may suffice.

Formulating a guideline for the use of generalizations,  \textit{generalizations shall be used to classify blocks, flowitems and ports by important terms and concepts used throughout the development}.
This concept especially comes into focus, when dealing with anomaly detection and failure detection, identification and recovery.
For example, the questions \ref{question_What_happens_if_component_X_breaks?} (concerning the fallout of a broken component) and \ref{question_What_happens_if_component_X_breaks?}b (analyzing components offline due to an electrical short in a component) would benefit from employing the following two concepts:
\begin{itemize}
  \item a \textit{fuse} class, which shows what fuse is the next upstream on the power path, that is triggered by the short and
  \item a  \textit{voltage} or \textit{power} class, which can be used to discern the power paths from data paths or other physical values for example.
\end{itemize}
In the same way, a \textit{data} class would be beneficial to answer Question \ref{question_What_happens_if_component_X_breaks?}c (concerning components suffering from a loss of input, as they depend on data processed by any of the components offline due to the electrical short in the a specified component), which depends on being able to discern a telemetry value from the physical flow it measures.

\subsubsection{Internal Block Diagrams and Modelling of Flowitems}

The purpose of Internal Block Diagrams is to show the connections between blocks and thereby define which paths flowitems can take within the model.

To answer questions such as Question \ref{question_What_component_supplies_system_X_with_power?} (querying the component that supplies a system with power), Question \ref{question_How_is_information_Y_being_processed_within_a_certain_subsystem?}  (concerning how a certain information is being processed within a certain subsystem)
Question \ref{question_global_datapath}, Question \ref{question_Which_components_draw_power_from_a_certain_supply_component?} (querying which components draw power from a certain supply component),
Question \ref{question_What_is_the_source_of_telemetry_Y_and_what_could_influence_its_measurement?} (querying the source of and possible influences on a certain telemetry value) or
Question \ref{question_Given_an_anomaly_on_a_specific_subset_of_a_system's_telemetry,__which_components_are_most_likely_to_have_caused_it?_Which_components_can_be_ruled_out?}
requires detailed information on the data and power flows within a system.
Therefore, \textit{Blocks used as flowitems should be modelled to the same level of detail with which analyses are to be performed later}.
To elaborate on this, the associations between flowitems on different levels, such as "Temperature X1  is part of the telemetry of Panel X, the telemetry of Panel X is part of Subsystem Y's telemetry" should be modelled on the same level of depth that is afterwards required for analyses.
As the analyses often start with anomalies seen on a single telemetry value this usually requires modelling down to the measurement of every single sensor.

While the flowitems should be modelled to great depth, the itemflows modelled in internal block diagrams do not require the same level of detail.
If a certain telemetry shall be traceable all the way back to the sensor generating it, the same level of detail also has to be provided in the internal block diagram.
Most modern spacecraft developments rely on commercial of the shelf parts to some degree.
The suppliers of these components typically do not provide the necessary information to model  on a level of detail allowing to track every data path within their subsystem.
Therefore,  the graph analysis has to be able to cope with black boxed subsystems, of course following the paradigm that you can only query information which is available in the model.
I.e.  if only the specification of the telemetry provided from the subsystem is available, but no information on its internal connections, setting the subsystem as a black box and transmitting a block containing all telemetry suffices.

\begin{figure}
  \centering
  \includegraphics{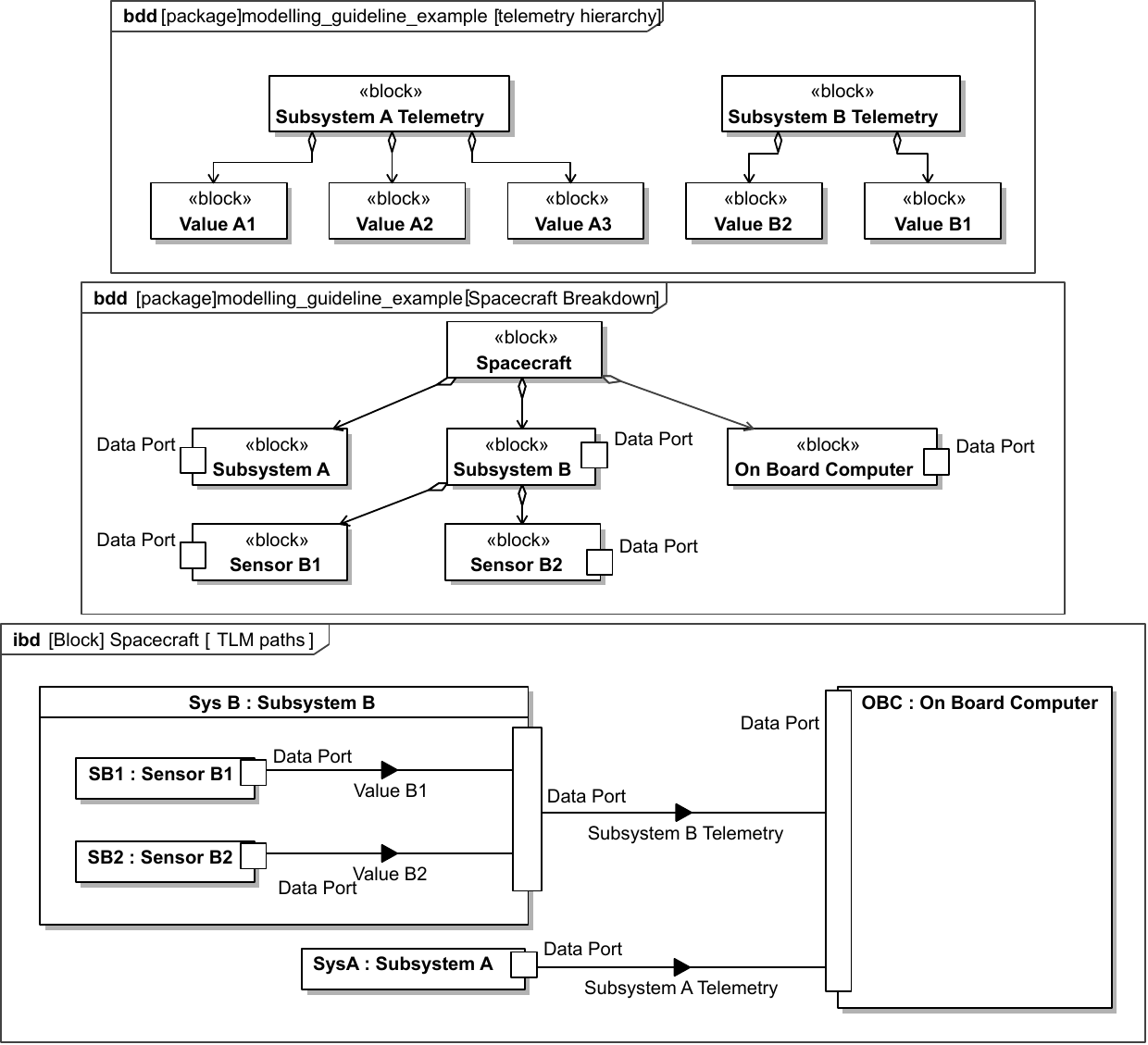}
  \caption{Exemplary application of the modelling guidelines for structural SysML aspects}
  \label{fig_ibd_and_flowitems_graph_schema_example}
\end{figure}

\Cref{fig_ibd_and_flowitems_graph_schema_example} illustrates this in an example.
The upper diagram in \Cref{fig_ibd_and_flowitems_graph_schema_example} shows the telemetry flowitems of \textit{Subsystem A} and \textit{Subsystem B}.
The middle diagram provides a structural breakdown of the spacecraft.
The bottom diagram shows the telemetry connections within the spacecraft in  the form of an internal block diagram.
Note here, that \textit{Subsystem A}, for  which no further breakdown is provided simply transmits its telemetry, which according to the upper diagram contains the values \textit{A1, A2} and \textit{A3}.

Depending on the system to be modelled, discerning between physical values and their measurements as flowitems can be a useful addition to the modelling guidelines.
An example of this would be an electrical power system, where currents, voltages and the measurements thereof are transmitted.

In such a case the application of generalizations of the flowitems to the two blocks \textit{physical value}  and \textit{measurement} resolves any ambiguity.
This is especially recommended to answer questions such as Question \ref{question_What_happens_if_component_X_breaks?} as it allows to backtrack power paths to the next fuse and to discern between power paths and data paths.
\FloatBarrier

\subsection{Modelling Guidelines for Behavioral Aspects of a SysML-Model}

Answering the questions set in \Cref{Analyzing the Behavioral Part of a SysML Model} also requires some modelling rules.
Compared to the structural part of SysML however, these are rather sparse.
Modelling state machine conditions with signals, as shown in Figures \ref{fig_simple_stm} and \ref{fig_simple_stm_trafo_v2} allows to answer Question \ref{question_shortest_path_states} a and c (What is the shortest path from state A to state B while condition C cannot be met? Which conditions have to be fulfilled for this path) and
Question \ref{question_condition_k_cannot_be_met_which_states_cannot_be_reached_anymore} (In case condition C cannot be met, which states of the system cannot be reached anymore).

Connecting state machines can be achieved via reusing state machines or employing signals.
Both of these practices are encouraged to achieve a higher level of interrelated, queryable information in the model.
The same concepts apply to Activity Diagrams.
When modelling activities, reusing existing activities to connect the diagrams can be of help to later retrieve information via the graph database.

\subsection{Summary}
Overall, the modelling guidelines to transfer SysML models to Neo4j in a way that enables such complex queries as described in \Cref{Analyses for a SysML Graph Schema} are few.
This was intended from the beginning, as the method described in this paper should be combinable with any other modeling strategy.
Parts of the rules described above can even be neglected, if the specific questions related to these rules are not of interest.


\section{Implementation}
\label{sec_implementation}

\begin{figure}
  \centering
  \includegraphics[width=.6\textwidth]{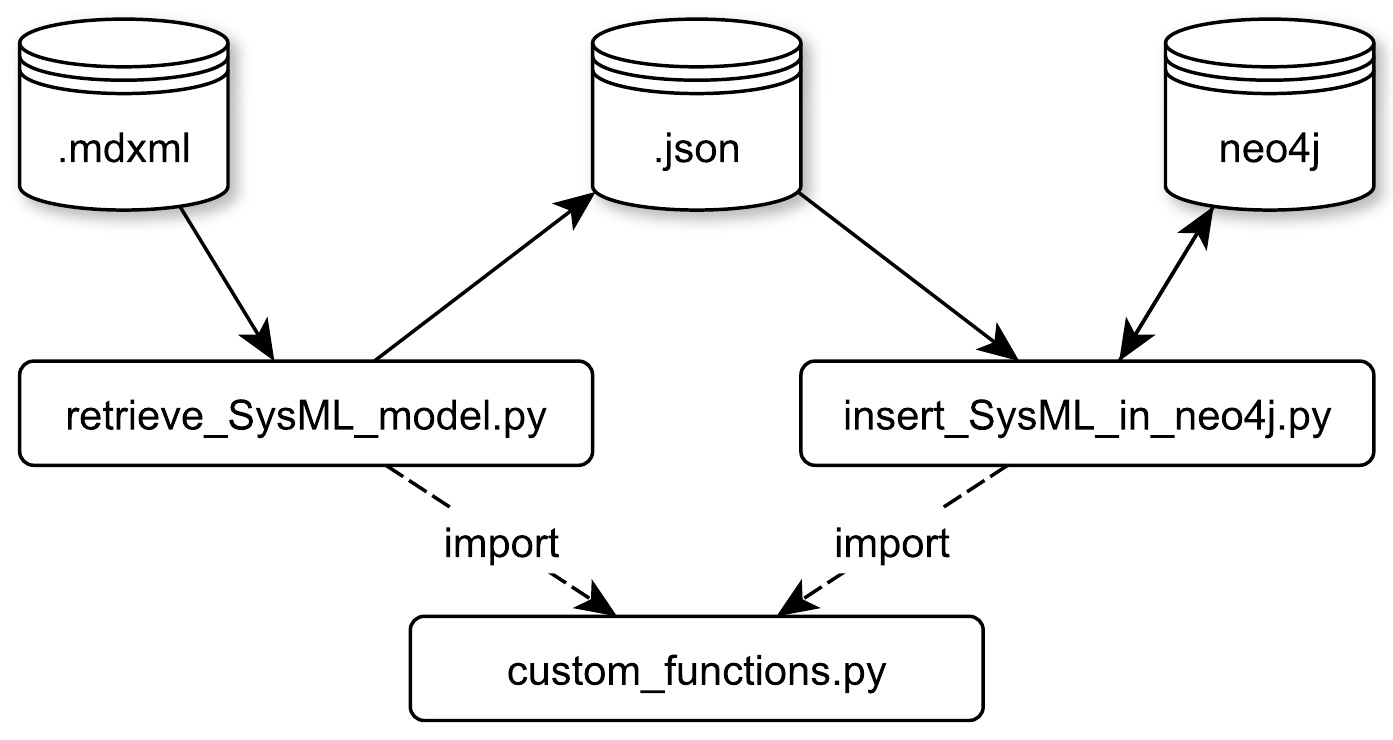}
  \caption{Setup of the extract transfer load software to import MagicDraw SysML data to Neo4j}
  \label{fig_implementation_schema}
\end{figure}

The translation of MagicDraw SysML Models into the graph is implemented using Python 3.8, with two scripts and a library of common functions, compare \Cref{fig_implementation_schema}.

\Cref{fig_implementation_schema} provides an overview of the implementation.
The \texttt{retrieve\_SysML\_model.py} script reads the SysML-Model generated in MagicDraw 19.0 (\texttt{.mdxml}-file) and extracts SysML components and relations as lists of python \texttt{dict} variables.
The lists are stored as \texttt{.json}-files.
This enables a separation of concerns and enables testing the retrieve functions separate from the insert functions, which insert the extracted SysML components in Neo4j.
It also enables future developers to use only the extraction part of our code and insert the extracted SysML components anywhere else.

The \texttt{insert\_SysML\_in\_neo4j.py} script loads the stored \texttt{.json} files and writes every component via a separate query into the Neo4j database.
This procedure is quite ineffective regarding execution times but allows for faster debugging.
As the execution time on a normal office PC is still under 3 minutes for several thousand SysML elements, the advantage in debugging prevails.

The code is open sourced on github under MIT License. It can be found under the following link:
\url{https://gitlab.lrz.de/lrt/sysml_graph_analysis_tool/}
\FloatBarrier

To help with initial studies of the subject, the reference models used in this paper, i.e. the MOVE-II Model and a smaller model containing the diagrams shown in \Cref{sec:Development of the Graph Schema} are published under \url{https://mediatum.ub.tum.de/1633734}

\FloatBarrier
\section{Application of the Schema on the MOVE-II Spacecraft}

\label{sec_application}

The following section describes the application of the schema and guidelines developed in \Cref{sec:Development of the Graph Schema} and \Cref{sec_modelling_guideline}.
After a short introduction, a selection of questions from \Cref{Analyzing the Structural Part of a SysML Model} are taken and respective queries are explained and performed on the SysML Model of the MOVE-II spacecraft.
To conduct these analyses, the SysML Model, which was generated using MagicDraw v19.0 SP4 was transformed into a Neo4j graph database, employing the schema built in \Cref{sec:Development of the Graph Schema} and the software implementation explained in \Cref{sec_implementation}.

The model of the MOVE-II spacecraft consists of a total of 605 blocks, 345 ports, 2820 relations, no activities and no states and thereby comprises a medium sized model with purely structural aspects.
No information on the behavior is included.

The model encompasses the satellite, ground station and operations systems and shows the data paths of all telemetry the spacecraft provides as well as all power paths within the spacecraft.
The spacecraft itself consists of multiple subsystems;
\begin{itemize}
  \item Attitude Determination and Control System, controlling the spacecraft's attitude relative to the Earth and Sun
  \item Communications (consisting of S-Band and UHF/VHF Transceiver), which provides contact to the ground
  \item Command and Data Handling, which handles all telementry, interprets communications and controls the state of the satellite
  \item Electrical Power System, containing the spacecraft's batteries and power converters and controlling the maximum power point
  \item Solar Cell Payload, solar cells which shall be measured against degradation over time in the space environment
  \item Structure and Mechanisms, providing structural integrity and deploying the antennas and solar array
\end{itemize}

\subsection{Hierarchical Analyses}
The same list of subsystems provided above can be generated from the graph database with the following cypher-query, which searches for a node with the label \texttt{:BLOCK} and \texttt{name}-property \textit{MOVE-II satellite} and any further block, which is directly a part of \textit{MOVE-II satellite}, here described with the \textit{subsystem}-variable.
Query \ref{query_retrieve_parts} returns the \texttt{name} property of all nodes that match the position of the \textit{subsystem}-variable:
\begin{listing}[!h]
  \caption{Retrieving the parts of a Block. \textit{Note:} By adding \texttt{*} after \texttt{IS\_PART\_OF} the query retrieves the parts to unlimited depth.}
  \label{query_retrieve_parts}
  \begin{minted}{cypher}
    MATCH (moveii:BLOCK{name:'MOVE-II satellite'}) <-[:IS_PART_OF]- (subsystem:BLOCK)
    RETURN subsystem.name
  \end{minted}
  \begin{minted}{cypher}
    subsystem.name
    ADCS
    CDH
    EPS
    UHF/VHF
    PL
    S-Band
    Solar Array
    STR
  \end{minted}
\end{listing}

While the above example already shows the solution for Question \ref{question_What_is_component_X_composed_of?}
from \Cref{Analyzing the Structural Part of a SysML Model} (What is component X composed of?),
the Query \ref{query_retrieve_port_types} answers Question \ref{question_What_types_of_data_ports_are_used_over_a_certain_range_of_equipment?} (What types of ports are used over a certain range of equipment?).
It starts by anchoring the query to the node with a \texttt{:BLOCK} label and the name \textit{Ground Station}, asks for any \texttt{:PORT} that \texttt{:IS\_PART\_OF} the ground station node and queries the port types via the \texttt{:IS\_OF\_TYPE} relations
\footnote{The asterix behind \texttt{:IS\_PART\_OF} defines that an arbitrary number of relations of the type can be followed in this direction.}.
It returns distinct names of the \textit{porttype} variables and their number of usages, ordering the results in descending order by the number of usages of the port type within the ground station. The result is provided below the query.
Note how the declarative definition of relation types fits the declarative language-style of Cypher, allowing for queries that can be understood with little prior knowledge of the language.
\begin{listing}[!h]
  \caption{Retrieve all port types used within a certain range of equipment.}
  \label{query_retrieve_port_types}
  \begin{minted}{cypher}
    MATCH (GroundStation:BLOCK{name:'Ground Station'}) <-[:IS_PART_OF*]- (PortInGs:PORT)-[:IS_OF_TYPE]->(porttype)
    RETURN DISTINCT porttype.name, count(porttype) ORDER BY count(porttype) DESC
  \end{minted}
  \begin{minted}{cypher}
    porttype.name	    count(porttype)
    'N Connector'	    8
    'Ethernet'    	   3
    'Serial Port'  	  2
    'USB'	            2
    'SMA Connector'	  2
    'data port'	      1
  \end{minted}
\end{listing}

\FloatBarrier

\subsection{Tracing Data Paths and Analyzing Data Anomalies}
\subsubsection{Datapath Query}
\begin{figure}[!]
  \centering
  \includegraphics[width=\textwidth]{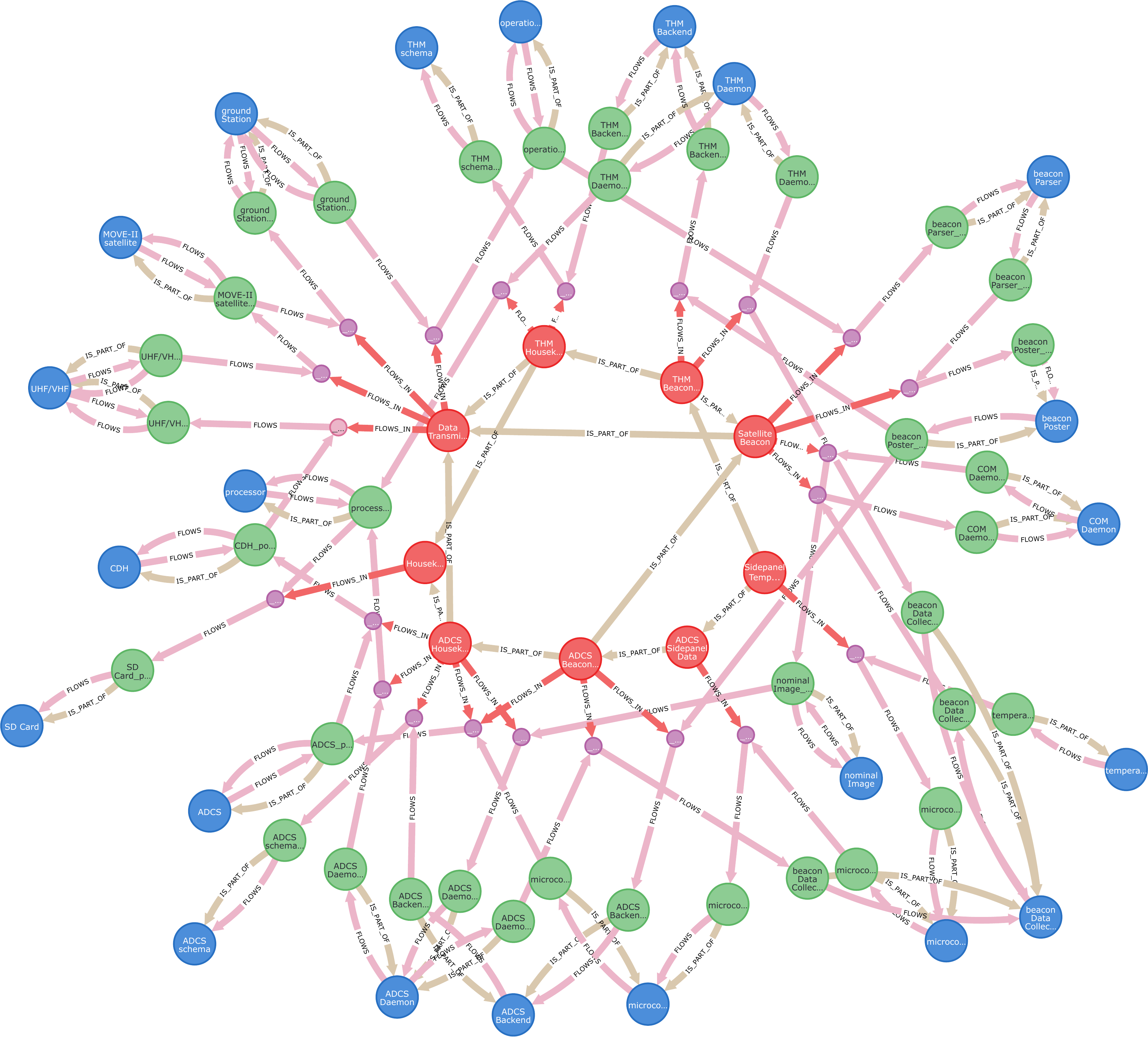}
  \caption{Graph representation of the data path query. Red: \texttt{:FLOWITEM} nodes, green: \texttt{:PORT} nodes, blue: \texttt{:BLOCK:INSTANCE} nodes, purple: \texttt{:HYPERNODE}s }
  \label{fig_datapath_query}
\end{figure}

One of the first steps in analyzing anomalies is finding all components which potentially participate in the anomaly.
Taking a data anomaly as example, any component albeit software or hardware processing the anomaly-holding telemetry are to be found.
Instead of consulting a variety of SysML Diagrams to find all components in question, which is cumbersome and prone to errors, Query \ref{query_datapath} can provide the result, defining any source and target component and the dataform in which the telemetry is being transmitted.
Note how the use of a parameter allows the reuse of the query for any other telemetry.
In this case, the telemetry is \textit{Sidepanel X+ Temperature OW2}, a temperature value on the outside of the spacecraft.
Query \ref{query_datapath} then looks for the \texttt{:FLOWITEM} with the defined telemetry name and all \texttt{:FLOWITEM}s which contain the node with the defined telemetry name.
In the next step, it finds all \texttt{:HYPERNODE}s where any of the \textit{flowitem} nodes \texttt{:FLOWS\_IN} and calls these \textit{hpn}.
The final step is to look for the pattern of \texttt{:BLOCK:INSTANCE} nodes, containing \texttt{:PORT}s connected via \texttt{:FLOWS} to a \textit{hpn}.
Over the \texttt{:FLOWS}-direction, the query differs between \textit{source} and \textit{target} of the flow and consequently returns a table with the results.
\begin{listing}[!h]
  \caption{Retrieve the datapath of a certain piece of information within the model. \textit{Note:} The response was cut short here and originally contains an additional 18 lines of response omittted for readability.}
  \label{query_datapath}
  \begin{minted}{cypher}
    :param searchterm=>'Sidepanel X+ Temperature OW2'
    //datapath table
    MATCH(telemetry:FLOWITEM {name: $searchterm}) -[:IS_PART_OF*0..]-> (flowitem:FLOWITEM)
    WITH flowitem
    MATCH(flowitem)-[:FLOWS_IN]->(hpn:HYPERNODE)
    WITH flowitem, hpn
    MATCH path = (source:BLOCK:INSTANCE) <-[:IS_PART_OF]- (:PORT) -[:FLOWS]-> (hpn) -[FLOWS]-> (:PORT) -[:IS_PART_OF]-> (target:BLOCK:INSTANCE)

    RETURN DISTINCT flowitem.name AS processedElement, source.name AS source, target.name AS target ORDER BY processedElement
  \end{minted}
  \footnotesize

  \begin{tabularx}{\textwidth}{l l l}
    \hline
    \texttt{processedElement}         & \texttt{source}                  &\texttt{target}
    \\
    \texttt{'ADCS Beacondata'}        & \texttt{'beacon Poster'}         & \texttt{'ADCS Backend'}  \\
    \texttt{'ADCS Beacondata'}        & \texttt{'ADCS Daemon'}           & \texttt{'beacon Data Collector'}  \\
    \texttt{'ADCS Beacondata'}        & \texttt{'microcontroller'}       & \texttt{'ADCS'}  \\
    \texttt{'ADCS Housekeeping Data'} & \texttt{'ADCS Backend'}	         &\texttt{'ADCS schema'}  \\
    \texttt{'ADCS Housekeeping Data'} & \texttt{'ADCS'}                  & \texttt{'CDH'}  \\
    \texttt{'Sidepanel X+ Temperature OW2'} & \texttt{'temperature Sensors x+'} & \texttt{'microcontroller x+'}  \\
    \texttt{[...]} & &\\\hline
  \end{tabularx}
\end{listing}
Of course, the same query can be performed to return a graph instead of a mere table, which might be useful to graphically assist the understanding of the dataflow.
\Cref{fig_datapath_query} shows the result of this query. Note how at a certain size the graph loses visual interpretability.

\subsubsection{Finding Probable Causes of Anomalies}
The next step in finding a data anomaly is to identify other telemetry processed by the components Query \Cref{query_datapath} yielded and checking them for anomalies.
A component often either processes any data correctly or none, so while this is just an empirical step, it is an important and useful one.
Query \ref{query_bad_anomaly_suggestions} is an addition to Query \ref{query_datapath} and yields any other telemetry processed by the same ports.
\begin{listing}[!h]
  \caption{Retrieve suggestions for possibly compromised telemetry  by checking telemetry which is directly processed by the same components.}
  \label{query_bad_anomaly_suggestions}
  \begin{minted}{cypher}
  CALL{
      MATCH (telemetry{name: $searchterm})-[:FLOWS_IN]->(hpn) RETURN hpn, telemetry
      UNION
      MATCH (telemetry{name: $searchterm})-[:FLOWS_IN]->()-[:FLOWS]-()-[:IS_PART_OF]->(component)<-[:IS_PART_OF]-()-[:FLOWS]-(hpn) RETURN hpn, telemetry
      }
  WITH hpn, telemetry
  MATCH (suggestion)-[:FLOWS_IN]->(hpn)
    WHERE NOT suggestion = telemetry
  return suggestion.name ORDER BY SHORTESTPATH((telemetry)-[:FLOWS_IN|FLOWS*]-(suggestion))
  \end{minted}
  \begin{minted}{cypher}
        suggestion.name
  	'Sidepanel X+ Temperature OW3'
  	'Sidepanel X+ Temperature OW1'
  	'PDM Current ADCS 3V3 1'
  	'ADCS Sidepanel Data Package x+'
  	'Sun Vector x+'
  	'Sidepanel X+ Temperature BMX'
  	'Gyroscope Data x+'
  	'Magnetic Field Vector x+'
  \end{minted}
\end{listing}

This query also builds the basis to answer Question \ref{question_Given_an_anomaly_on_a_specific_subset_of_a_system's_telemetry,__which_components_are_most_likely_to_have_caused_it?_Which_components_can_be_ruled_out?}
(Given an anomaly on a specific subset of a system's telemetry,  which components are most likely to have caused it? Which components can be ruled out?).

The question refers to a scenario, where more than one telemetry value shows an anomaly.
To rule out any other component, the query follows the logic "if another input is being processed by the exact same component and port correctly, the error is most likely not in this component."
Query \ref{query_list_of_fault_candidates_for_tlm_fault_pattern} shows the basic pattern to find components processing the three faulty telemetry packages \textit{\$fltm1, \$fltm2, \$fltm3}, while not processing the healthy telemetry package \textit{\$good\_tlm}.
The result shows a list of ports and components, including their IDs.
The query shows how it is possible to narrow down a list of over 1400 possible suggestions to a mere 6 by applying logic on the graph transformation of the SysML model.
Taking a closer look at the proposed components, we find that the ports \textit{p3} and \textit{p4} are proxy ports and therefore no real components.
Ruling those out, we end up with one component and three ports as suggested causes of the anomaly.
It has to be noted here, that other components or ports could be at fault as well, as not every possible fault path is traceable through the model.
However the query provides a good starting point for the analysis.

\begin{figure}
  \centering
  \includegraphics[width=\textwidth]{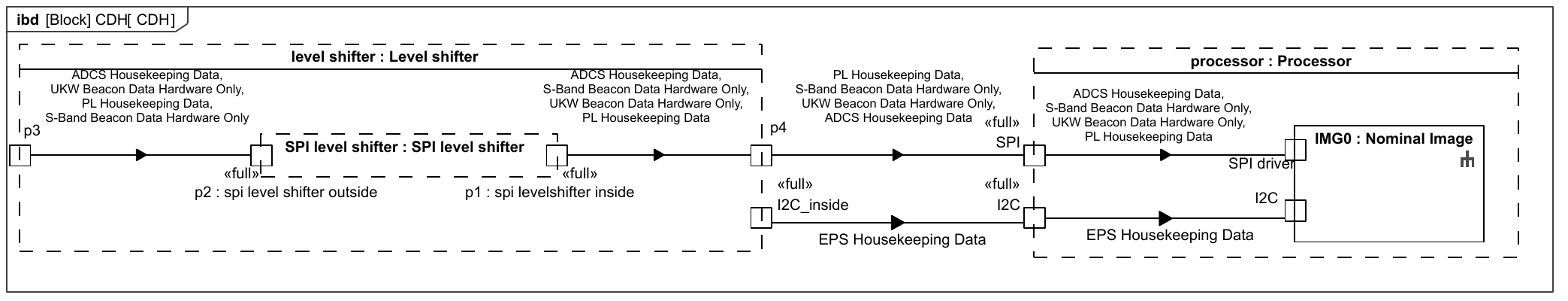}
  \caption{Internal Block Diagram to Query \ref{query_list_of_fault_candidates_for_tlm_fault_pattern}}
  \label{figure_ibd_faulty_tlm}
\end{figure}

\Cref{figure_ibd_faulty_tlm} shows the respective part of the SysML Model, including the item flows, ports and blocks.

\begin{listing}[!h]
  \caption{Determine any port or block processing a list of faulty telemetry while not processing healthy telemetry.}
  \label{query_list_of_fault_candidates_for_tlm_fault_pattern}

  \begin{minted}{cypher}
    :param fltm1=>'ADCS Housekeeping Data'
    :param fltm2=>'UKW Beacon Data Hardware Only'
    :param fltm3=>'S-Band Beacon Data Hardware Only'
    MATCH (faulty_tlm1{name:$fltm1}), (faulty_tlm2{name:$fltm2}), (faulty_tlm3{name:$fltm3}), (good_tlm {name: $good_tlm})
    WITH *
    MATCH (faulty_tlm1)-[:FLOWS_IN]->()-[:FLOWS]->()-[:IS_PART_OF]->(component)
    WHERE EXISTS {(faulty_tlm_2)-[:FLOWS_IN]->()-[:FLOWS]->()-[:IS_PART_OF]->(component)}
        AND EXISTS {(faulty_tlm3)-[:FLOWS_IN]->()-[:FLOWS]->()-[:IS_PART_OF]->(component)}
        AND NOT EXISTS{(good_tlm)-[:FLOWS_IN]->()-[:FLOWS]->()-[:IS_PART_OF]->(component)}
    RETURN component.name as suggestion, component.id as id
    UNION
    MATCH (faulty_tlm1{name:$fltm1}), (faulty_tlm2{name:$fltm2}), (faulty_tlm3{name:$fltm3}), (good_tlm {name: $good_tlm})
    WITH *
    MATCH (faulty_tlm1)-[:FLOWS_IN]->()-[:FLOWS]->(port)
    WHERE EXISTS {(faulty_tlm_2)-[:FLOWS_IN]->()-[:FLOWS]->(port)}
        AND EXISTS {(faulty_tlm3)-[:FLOWS_IN]->()-[:FLOWS]->(port)}
        AND NOT EXISTS {(good_tlm)-[:FLOWS_IN]->()-[:FLOWS]->(port)}
    RETURN port.name as suggestion, port.id as id
  \end{minted}
  \footnotesize

  \begin{tabularx}{\textwidth}{l X }
    \hline
    \texttt{suggestion}	& \texttt{id}\\
    \texttt{SPI level shifter}	& \texttt{\_19\_0\_4\_9b3028f\_1626770558674\_761066\_49622}\\
    \texttt{CDH\_SPI\_port\_instance}	& \texttt{\_19\_0\_4\_64d021d\_1606812551466\_659190\_43643 \_19\_0\_4\_9b3028f\_1626784294211\_953036\_61690}\\
    \texttt{processor\_port\_instance}	& \texttt{\_19\_0\_4\_9b3028f\_1606755365738\_580408\_43811 \_19\_0\_4\_9b3028f\_1626770585221\_986561\_49783}\\
    \texttt{level shifter\_p4\_port\_instance}	& \texttt{\_19\_0\_4\_9b3028f\_1632823793184\_527010\_45978 \_19\_0\_4\_9b3028f\_1626770580938\_1308\_49739}\\
    \texttt{level shifter\_p3\_port\_instance}	& \texttt{\_19\_0\_4\_9b3028f\_1632823808903\_57539\_46012 \_19\_0\_4\_9b3028f\_1626770580938\_1308\_49739}\\
    \texttt{SPI level shifter\_port\_instance}	& \texttt{\_19\_0\_4\_64d021d\_1609248717945\_193014\_44390 \_19\_0\_4\_9b3028f\_1626770558674\_761066\_49622}\\
    \hline
  \end{tabularx}
\end{listing}

\FloatBarrier
\subsection{Failure Propagation Analyses}
The next type of analysis already becomes useful during the design of the spacecraft.
In comparison to ground-based systems, robustness against failure is a design goal for any spacecraft, as on-sight repairs are usually not possible.
Hence failure propagation is to be kept to an absolute minimum, i.e. the spacecraft's attitude control system, on board computer, power and basic communication systems should not be influenced by a failure in any other part of the system.

The  graph schema and modelling guidelines proposed here do not enable the analysis of complex relations such as "the communication system can only work if the spacecraft is pointing to the ground station".
The analyses possible with the here proposed graph schema and modelling guidelines center around Question \ref{question_What_happens_if_component_X_breaks?} (What happens if component X breaks?).

In case the component X is a physical component, a broken component may trigger an electrical short.
To prevent a spread of such an event electrical fuses are employed throughout the system.

Therefore one of the tasks to be applied here is to find out which fuse triggers and which components are shut down as well by the fuse.
Query \ref{query_shorted_components} does exactly this.
However, to enable this query, the graph needs to extend by an additional \texttt{:FLOWS} relation between any \texttt{:PORT} with an incoming or outgoing flow connection and the \texttt{:BLOCK} which the port is part of.
This is accomplished by the code presented in Query \ref{query_add_flow_between_ports_and_blocks}.
Note further that Query \ref{query_shorted_components} takes two parameters as input; the \textit{\$searchterm} parameter defines which component was shorted, the \textit{\$flowlength} parameter defines how many \texttt{:FLOWS} connections the query shall follow before it stops searching.
It also requires to follow the modelling schema presented in \Cref{sec_generalizations_modelling_guideline}, defining \textit{physical flowitems} as a class differing from telemetry in order to enable an accurate tracking of the power path.
Limiting the amount of \texttt{:FLOWS} connections followed in the query is not technically necessary but increases the performance significantly.
As the volatile relations were all built with the property \textit{tbd=True}, they can be deleted again after the query is finished.

\begin{listing}[!h]
  \caption{Code to create volatile [:FLOWS] relations necessary to apply the shortest path algorithm in Query \ref{query_shorted_components}.}
  \label{query_add_flow_between_ports_and_blocks}
  \begin{minted}{cypher}
    MATCH (b:INSTANCE:BLOCK)<-[:IS_PART_OF]-(p:PORT)-[:FLOWS]->(:HYPERNODE)
    MERGE (p)<-[volatile_rel_out:FLOWS]-(b)
    SET volatile_rel_out.tbd = True

    MATCH (b2:INSTANCE:BLOCK)<-[:IS_PART_OF]-(p2:PORT)<-[:FLOWS]-(:HYPERNODE)
    MERGE (p2)-[volatile_rel_in:FLOWS]->(b2)
    SET volatile_rel_in.tbd = True
  \end{minted}
\end{listing}
\begin{listing}[!h]
  \caption{Find components affected by an electrical short.}
  \label{query_shorted_components}
  \begin{minted}{cypher}
    Call {
      MATCH p = (rtc{name:$searchterm}) <-[:IS_PART_OF]- (port) <-[:FLOWS]-()<-[:FLOWS_IN]- (shortedvoltage) -[:IS_OF_TYPE]-> (pfi{name:'physical flowitem'})
      WITH shortedvoltage
      MATCH (fusetype {name:'electrical fuse'}) <-[:IS_OF_TYPE]- (fuse) <-[:IS_INSTANCE_OF]- (fuseinstance) <-[:IS_PART_OF]- (fuseportinstance) -[:FLOWS]-> (hpn) <-[:FLOWS_IN]- (shortedvoltage)
      RETURN fuseinstance, shortedvoltage ORDER BY length(SHORTESTPATH((fuseinstance) -[:FLOWS*]- (rtc))) LIMIT 1}
    WITH *
    MATCH p=(fuseinstance) -[:FLOWS*1..$flowlength]-> (hpn:HYPERNODE) <-[:FLOWS_IN]- (shortedvoltage)
    WITH hpn, fuseinstance
    MATCH (hpn) -[:FLOWS]-> () -[:IS_PART_OF]-> (shortedcomponent)
    WHERE NOT fuseinstance.id = shortedcomponent.id
    RETURN shortedcomponent.name
  \end{minted}
  \begin{minted}{cypher}
    shortedcomponent.name
    'current Limiter SD Cards'
    'SD Card'
    'FRAM'
    'flash'
    'GPS'
    'processor'
    'real Time Clock'
    'level shifter'
  \end{minted}
\end{listing}

Query \ref{query_shorted_components} can also be tuned to show any telemetry values that are not processed correctly anymore due to the short by adding a new \texttt{MATCH}-clause based on the variable \textit{shortedcomponent}.

\subsection{Summary}

As shown on select examples above, the graph allows to retrieve information from a complex context in a minimum of time.
Especially noteworthy are the efficiency increase on looking for similar information within a different context and the ability to query complex design constructs.
The parametrization of the queries allows to reuse them with different objects of interest and even on different systems.
Some of the queries here would benefit from a more compact graph schema.
Setting up the complete graph in a simpler schema inevitably results in a loss of information, though.
The graph projection capability of Neo4j however allows to create simpler projected graphs from an existing graph, which could be used to shorten queries (see \cite{neo4jgdsmanual}), without losing information.

\FloatBarrier
\section{Conclusion}
\label{sec_conclusion}

Over the past four sections of this paper, a graph schema and modelling guidelines evolved from a set of analysis questions which are typical for assembly, integration and testing of small spacecraft and are based on personal experience.
\Cref{sec:Development of the Graph Schema} developed the graph schema as a mere concept, weighing different options and explaining the thought process behind the proposed schema.
To the best of the author's knowledge it is the only public schema for SysML interpretation in graph databases to this date that goes beyond the modelling of requirements and use cases presented by \cite{petnga2019}.
In the form proposed here it completely lacks any schema for the handling of use cases, requirements, sequence diagrams and parameter diagrams as well as custom stereotypes.
Seeing as this paper is already quite lengthy this was not an oversight but the result of a pragmatic decision.
\Cref{sec_modelling_guideline} interlaces the defined graph schema with the questions defined in \Cref{Analyses for a SysML Graph Schema} into a concise set of modelling rules, while explaining where modellers can cut corners on the rules depending on their analysis goals.
\Cref{sec_implementation} lays out the code structure on a basic level and provides the reader with pointers to the respective open source repository and the two published test models.
We hope the code provided in the repository enables any interested party to apply the schema at their own models.
\Cref{sec_application} shows the application of the schema on the model of the MOVE-II spacecraft.
Queries are deduced from selected questions defined in \Cref{Analyses for a SysML Graph Schema} to show the capabilities and limitations of schema and modelling guidelines.
The queries go far beyond the publications by \cite{petnga2019},\cite{fisher20143,bajaj2016mbse++} and\cite{bajaj2017graph}.
The clear focus on the information a graph shall provide and the parallel development of modelling rules enable far reaching and complex analyses on the system.

Overall the paper shows how graph analyses based on SysML can be a useful tool for engineering teams working with MBSE.
It takes on the challenges to the adoption of MBSE summarized in \Cref{sec:stateoftheart}, especially regarding the insufficiency of tools, the low share of projects alraedy implementing MBSE and the importance of keeping with prevalent modelling strategies (compare \cite{nasa-se-study2019}),  while keeping the functional complexity in control and structuring the knowledge about the system in a traceable manner (cf. \cite{mb4se-2020-userneeds}).

A weak point of the provided implementation is its focus on files created with MagicDraw v19.0 SP4.
As we do not have access to other versions of the modelling software, no cross tests with newer software versions was possible.
A second weak point is the lack of proposed queries with regard to the behavioral aspects of SysML.
This is due to a lack of a behavior-intensive medium-scale SysML model on the one side and on the other side due to a trade-off between explaining more challenging and complex queries on the structural part of the model and the length of the text.

\section{Outlook}
\label{sec_outlook}

While the above presented schema and application contain a solid proposal for
graph transformations of information in Block Definition Diagarms, Internal Block Diagrams, Activity Diagrams and State Machines, these make only four of the  9 diagram types of SysML  (cf. \cite[p. 211]{sysml-16}).
Schemata for Package-, Parametric-, Sequence-, Use Case- and Requirement Diagrams are still to be defined.
It furthermore does not cover the treatment of custom profiles which may require additional node and relation types.
All of this work may be done in future revisions of the schema and may be the topic of further publications.
Another interesting aspect would enhancing  formal reviews of a SysML model via graph queries.
Parametrized queries could be set to any model to be reviewed, increasing efficiency and transparency of the review process.
The comparison between different versions of a SysML Model via graph analysis could also be interesting and would most likely require some slight adaptions to the schema proposed here.
Only a small selection of the questions posed in \Cref{Analyses for a SysML Graph Schema} could be shown here.
Formulating the queries to answer all questions provided in \Cref{Analyses for a SysML Graph Schema} is a prospect for the future.

The graph database could be extended to include document-based information and relating it on a key-word basis to the SysML Model, thereby creating a holistic view of the system's digital environment.
First tests show that while a promising amount of relations can be created by such an approach, the processing of synonyms, abbreviations and unrelated usage of the same words create a unique challenge.
Compared to classical MBSE approaches, which require a high level of discipline from everyone involved, such an approach could be conducted without requiring any change of behavior by the engineering team, which - as stated by \cite{nasa-se-study2019,serc2020} and \cite{chamimbsesurvey} - could result in an easier introduction of MBSE in general.

\newpage

\begin{Backmatter}

\bibliographystyle{apalike}
\bibliography{main}

\end{Backmatter}

\end{document}